\documentclass{elsart}
\usepackage{natbib}
\usepackage{amssymb}
\usepackage{graphicx,amsmath}
\usepackage{epstopdf}
\usepackage{color}

\def\om{\omega}

\def\x{{\rm\bf x}}
\def\y{{\rm\bf y}}

\def\k{{\rm\bf k}}
\newcommand{\be}{\begin{equation}}
\newcommand{\ee}{\end{equation}}
\newcommand{\beq}{\begin{equation}}
\newcommand{\eeq}{\end{equation}}
\newcommand{\beqa}{\begin{eqnarray}}
\newcommand{\eeqa}{\end{eqnarray}}
\newcommand{\bq}{\begin{eqnarray}}
\newcommand{\eq}{\end{eqnarray}}

\newcommand{\ra}{\rangle}
\newcommand{\la}{\langle}
\newcommand{\beqax}{\begin{eqnarray*}}
\newcommand{\eeqax}{\end{eqnarray*}}
\newcommand{\beqas}{\begin{eqnarray*}}
\newcommand{\eeqas}{\end{eqnarray*}}

\newcommand{\pe}{\perp}
\newcommand{\pa}{\parallel}

\newcommand {\fcos} [1] {\cos \left( #1 \right)}
\newcommand {\fsin} [1] {\sin \left( #1 \right)}
\newcommand {\fabs}[1] {\left| #1 \right|}

\journal{Advances in Atomic, Molecular and Optical Physics}
\begin{document}

\begin{frontmatter}



\title{Shortcuts to adiabaticity}


\author[UPV]{E. Torrontegui}
\author[UPV]{S. Ib\'a\~nez}
\author[UPV]{S. Mart\'\i nez-Garaot}
\author[FT,IK]{M. Modugno}
\author[LA1,LA2]{A. del Campo}
\author[Toul]{D. Gu\'ery-Odelin}
\author[Cork]{A. Ruschhaupt}
\author[UPV,Shan]{Xi Chen}
\author[UPV,Shan]{J. G. Muga}


\address[UPV]{Departamento de Qu\'{\i}mica F\'{\i}sica, Universidad del Pa\'{\i}s Vasco - Euskal Herriko Unibertsitatea, 
Apdo. 644, Bilbao, Spain}
\address[FT]{Departamento de F\'{\i}sica Te\'orica e Historia de la Ciencia, Universidad del Pa\'{\i}s Vasco - Euskal Herriko Unibertsitatea,  Apdo. 644, Bilbao, Spain}
\address[IK]{IKERBASQUE, Basque Foundation for Science, 
48011 Bilbao, Spain}
\address[LA1]{Theoretical Division, Los Alamos National Laboratory, Los Alamos, NM, USA}
\address[LA2]{Center for Nonlinear Studies, Los Alamos National Laboratory, Los Alamos, NM, USA}
\address[Toul]{Laboratoire Collisions Agr\'egats R\'eactivit\'e, CNRS UMR 5589, IRSAMC, Universit\'e Paul Sabatier, 
31062 Toulouse CEDEX 4, France}
 \address[Cork]{Department of Physics, University College Cork, Cork, Ireland}
\address[Shan]{Department of Physics, Shanghai University, 200444 Shanghai, People's Republic of China}

\begin{abstract}

Quantum adiabatic processes --that keep constant the populations in the instantaneous eigenbasis of a time-dependent Hamiltonian-- are very useful to prepare and manipulate states,
but take typically a long time. This is        
often problematic because decoherence and noise may spoil the desired final state, or because some applications require many repetitions.  ``Shortcuts to adiabaticity'' 
are alternative fast processes which reproduce the same final populations, or even the same final state, as the adiabatic process in a finite, shorter time. Since adiabatic processes are ubiquitous, the shortcuts span a broad range of applications in atomic, molecular and optical physics, such as fast transport of ions or neutral atoms, internal population control and state preparation (for nuclear magnetic resonance or quantum information), cold atom expansions and other manipulations, cooling cycles, wavepacket splitting, and many-body state engineering or correlations microscopy. Shortcuts are also 
relevant to clarify fundamental questions such as a precise quantification of the third principle of thermodynamics
and  quantum speed limits.     
We review different theoretical techniques proposed to engineer the shortcuts, the experimental results, and the prospects.    
\end{abstract}


\end{frontmatter}

\tableofcontents
%
%
%
%
%
%
%
\section{Introduction}
The expression ``shortcuts to adiabaticity'' (STA) was recently 
introduced  in \cite{Ch10}, to describe protocols that speed up a quantum adiabatic process, usually, although not necessarily, through a non-adiabatic route.\footnote{ 
The word ``adiabatic'' may have two 
different meanings: the thermodynamical one (no heat transfer between system and environment) and the quantum one, as stated by \citet{BF28} in the adiabatic theorem: ``a physical system remains in its instantaneous eigenstate when a given perturbation is acting on it slowly enough and if there is a gap between the eigenvalue and the rest of the Hamiltonian's spectrum''. 
Here we shall always understand ``adiabatic'' in the quantum mechanical sense.}    
{There the} Lewis-Riesenfeld invariants were used to inverse engineer the time dependence of a harmonic oscillator frequency
between predetermined initial and final values so as to avoid final excitations. That paper and its companion on Bose-Einstein condensates \citep{MCRG09} have indeed triggered a surge of activity, 
not only for harmonic expansions \citep{energy,Muga10,Li10,Nice10,Nice11,delcampo11a,Nice11b,Li11,3d,Yidun,ErikFF,DB12,Li12}, but for atom transport \citep{transport,BECtransport,OCTtrans,Bowler}, quantum computing \citep{PLA2011}, quantum simulations \citep{simu},  optical lattice expansions \citep{PLA2012,Yuce2}, wavepacket splitting \citep{split}, internal state control \citep{Chen11,NHSara,noise,Ban,Sara12,Yidun1}, many-body state engineering \citep{delcampo11,DB12,DRZ12,JJ}, and other applications  such as sympathetic cooling of atomic mixtures \citep{Onofrio11,Onofrio12}, or 
cooling of nanomechanical resonators \citep{Lianao,Lianao2}. 
In fact several works had previously or simultaneously considered to speed up adiabatic processes making use of  different techniques. For example  \citet{Rice03,Rice05,Rice08} and \citet{Berry09} proposed the  
addition of counterdiabatic terms to a reference Hamiltonian $H_0$ to achieve  adiabatic dynamics with respect to $H_0$. This ``transitionless tracking algorithm'' 
\citep{Berry09} has been applied to manipulate the populations of two-level systems \citep{Rice05,Berry09,Ch10b,Oliver,expcd}. Another technique  
to design laser pulses for fast population transfer is parallel adiabatic passage \citep{PLAP1,PLAP2,PLAP3,PLAP4}. \citet{David} {designed 
trap motions in order} to perform non-adiabatic fast transport of atomic cold clouds. Also,  \citet{MNProc}
developed a ``fast-forward technique'' for several manipulations on wavepackets such as expansions, transport or splitting of Bose-Einstein condensates.  
Related work had also been carried out for wave packet splitting making use of optimal control \citep{S07,S09a,S09b}, and in the context of quantum refrigerators, to find fast ``frictionless'' expansions \citep{Salamon09,KoEPL09}. For recent developments on this line stimulated by invariant-based engineering results see \citet{KoPRE10,EPL11,KoPRL12,KoPRE12}. 

The considerable number of publications on the subject, and a recent Conference on ``Shortcuts to adiabaticity'' held in Bilbao (16-20 July 2012)
demonstrate much current interest, not only within  the cold atoms and atomic physics communities but also from fields such as semiconductor physics and spintronics \citep{Ban}. Indeed adiabatic processes are ubiquitous, so
we may expect a broad range of applications, even beyond the quantum domain, since some of the concepts are easy to translate into optics \citep{SHAPEapp,Tseng} 
or mechanics \citep{NHSara}.   
Apart from the practical applications,
the fundamental implications of shortcuts 
on 
quantum speed limits
\citep{qb,Oliver,Moise2012}, time-energy uncertainty relations \citep{energy},  multiple Schr\"odinger pictures \citep{Sara12}, and {the quantification of}  the third principle of thermodynamics
and of maximal cooling rates \citep{Salamon09,KoEPL09,energy,KoPRE10,EPL11,KoPRL12,KoPRE12} are also intriguing
and provide further motivation. 
 
In this review we shall first describe different approaches to STA
in Sec. \ref{gf}. While the main goal there is
to construct new protocols for a fast manipulation
of quantum states avoiding final excitations, 
additional conditions may be imposed. 
For example, ideally these protocols should not be state specific but work
for an arbitrary state\footnote{Contrast this to the quantum brachistochrone \citep{qb}, in which the aim is to 
find a time-independent Hamiltonian that takes a given initial state to a given final state in minimal time. Studies of ``quantum speed limits'' adopt in general this state-to-state approach, as in  \citet{Moise2012}.}. 
They should also be stable against perturbations, 
and keep the values of the transient energy and other variables 
manageable throughout the {whole} process.   
{Several} applications are discussed in Secs. 3 to 6. 
We have  kept {a notation consistency} within each Section but not 
throughout the whole review, following when possible notations close to the original publications.
%
%
%
%
%
%
\section{General Formalisms\label{gf}}
\subsection{Invariant Based Inverse Engineering}
\label{inveng}
{\it Lewis-Riesenfeld invariants.--}
The \citet{LR} theory
is applicable to a quantum system that evolves with a
time-dependent Hermitian Hamiltonian $H(t)$, which supports
a Hermitian dynamical invariant $I(t)$ {satisfying}
\beq
i \hbar \frac{\partial I(t)}{\partial t } - [H(t), I(t)]=0.
\label{invad}
\eeq
{Therefore} its expectation values 
for an arbitrary solution of the
time-dependent Schr\"{o}dinger equation
$i \hbar \frac{\partial}{ \partial t } |\Psi (t)\ra = H(t) |\Psi (t)\ra$,
do not depend on time. 
$I(t)$ can be used to expand $|\Psi(t)\ra$
as a superposition of ``dynamical modes'' $|\psi_n (t) \rangle$,
\beq
|\Psi (t)\ra=\sum_n c_n  |\psi_n (t) \rangle,\;\;
|\psi_n(t)\ra=e^{i\alpha_n(t)}|\phi_n(t)\ra,
\label{expan1}
\eeq
where $n=0,1,...$; $c_n$ are time-independent amplitudes,
and $|\phi_n (t)\rangle$ are orthonormal eigenvectors of the invariant $I(t)$,
\beq
\label{invariant}
I(t) = \sum_n   |\phi_n (t) \rangle \lambda_n\langle \phi_n (t)|.
\eeq
%
The $\lambda_n$ are real constants,
and the Lewis-Riesenfeld phases are defined as \citep{LR}
\beq
\label{LRphase}
\alpha_n (t) = \frac{1}{\hbar} \int_0^t \Big\langle \phi_n (t') \Big|
i \hbar \frac{\partial }{ \partial t'} - H(t') \Big| \phi_n (t')  \Big\rangle d t'.
\eeq
We use for simplicity a notation for a discrete spectrum of $I(t)$ but the generalization
to a continuum or mixed spectrum is straightforward.
We also assume a non-degenerate spectrum.
%
Non-Hermitian invariants and Hamiltonians have been considered for example in \citet{Gao91,Gao92,Lohe,NHSara}.

{\it Inverse engineering.--}
Suppose that we want to drive the system
from an initial Hamiltonian $H(0)$ to a final one $H(t_f)$,
in such a way that the populations in the initial and final instantaneous bases are the same, but admitting
transitions at intermediate times. To inverse engineer a time-dependent Hamiltonian $H(t)$ and achieve this goal, we
may first define the invariant through its eigenvalues and eigenvectors. 
The Lewis-Riesenfeld phases $\alpha_n(t)$ may also be chosen as arbitrary functions
to write down the
time-dependent unitary evolution operator ${U}$, 
%
\beq
{U}= \sum_n e^{i \alpha_n (t)} |\phi_n (t) \rangle \langle \phi_n (0)|.
\eeq
$U$ obeys 
$
i \hbar \dot{U}  = H(t) {U},
$
where the dot means time derivative. 
Solving  formally this equation for $H(t)=i\hbar \dot{U}{U}^{\dag}$,
we get
\beq
\label{inHa}
H(t)=  - \hbar \sum_n |\phi_n(t)\ra \dot{\alpha}_n  \la \phi_n(t)|
+ i \hbar \sum_n | \partial_t \phi_n (t) \rangle \langle \phi_n (t)|.
\eeq
According to Eq. (\ref{inHa}), for a given invariant there are many possible
Hamiltonians corresponding to different choices of
phase functions $\alpha_n(t)$.
In general $I(0)$ does not commute with $H(0)$,
so the eigenstates of $I(0)$, $|\phi_n (0) \rangle$, do not coincide with the eigenstates of $H(0)$. $H(t_f)$ does not
necessarily commute with  $I(t_f)$ either.
If we impose $[I(0), H(0)]=0$ and $[I(t_f), H(t_f)]=0$,
the eigenstates will coincide, which 
guarantees a state transfer without final excitations.
In typical applications the Hamiltonians $H(0)$ and $H(t_f)$ are given, 
and set the initial and final configurations of the external parameters. 
Then we define $I(t)$ and its eigenvectors accordingly, so that 
the commutation relations are obeyed at the boundary times and, finally, $H(t)$ is designed via Eq. (\ref{inHa}).
While the $\alpha_n(t)$ may be taken as fully free time-dependent phases in principle, 
they may also be constrained by a pre-imposed or
assumed structure of $H(t)$. 
Secs. 3, 4 and 5 {present} examples of how this works for expansions, transport and internal state control.

A generalization of this inverse method for non-Hermitian Hamiltonians was considered in \citet{NHSara}. 
Inverse engineering was applied  
to accelerate the slow expansion of a classical particle in a time-dependent harmonic oscillator without final excitation. This system may be treated formally as a quantum two-level system with non-Hermitian Hamiltonian \citep{Gao91,Gao92}.  

{\it Quadratic in momentum invariants.--}
\citet{LR} 
paid special attention to the time-dependent harmonic oscillator and its invariants quadratic in position and momentum. 
Later on  \citet{LL} found, in the framework of classical mechanics, the general form of the Hamiltonian compatible with  quadratic-in-momentum invariants, which includes non harmonic potentials. This work, and the corresponding quantum results  of \citet{DL} constitutes the basis 
of this subsection. 

A one-dimensional Hamiltonian with a quadratic-in-momentum invariant must have the form $H=p^2/2m+V(q,t)$,\footnote{$q$ and $p$ may denote operators or numbers.  The context should clarify their exact meaning.}
with the potential \citep{LL,DL} 
\beq
\label{Vinv}
V(q,t)=
-F(t)q+\frac{m}{2}\omega^2(t)q^2+\frac{1}{\rho(t)^2}U\left[\frac{q-q_c(t)}{\rho(t)}\right].
\eeq
$\rho$, $q_c$, $\omega$, and $F$ are arbitrary functions of time
that
satisfy the auxiliary equations
\beqa
\ddot{\rho}+\omega^2(t)\rho&=&\frac{\omega_0^2}{\rho^3},
\label{Erma}
\\
\ddot{q}_c+\omega^2(t)q_c &=& F(t)/m,
\label{alphaeq}
\eeqa
where $\omega_0$ is a constant.
Their physical interpretation will be explained
below and depends on the  operation.  
A quadratic-in-$p$ dynamical invariant is given, up to a constant factor, by
\beq
\label{invaq}
I=\frac{1}{2m}[\rho(p-m\dot{q}_c)-m\dot{\rho}(q-q_c)]^2
+\frac{1}{2}m\omega_0^2\left(\frac{q-q_c}{\rho}\right)^2
+U\left(\frac{q-q_c}{\rho}\right).
\eeq
Now  $\alpha_n$ in Eq. (\ref{LRphase}) satisfies \citep{LR,DL}
\beqa
\alpha_n =-\frac{1}{\hbar}\int_0^t  {\rm d} t'\left(\frac{\lambda_n}{\rho^2}+
\frac{m(\dot{q}_c\rho-q_c\dot{\rho})^2}{2\rho^2}\right),
\eeqa
and the function $\phi_n$ can be written as \citep{DL}
\beq\label{psin}
\phi_n (q,t)=e^{\frac{im}{\hbar}\left[\dot{\rho} q^2/2\rho+(\dot{q}_c\rho-q_c\dot{\rho})q/\rho\right]}\frac{1}{\rho^{1/2}}\Phi_n\bigg(\underbrace{\frac{q-q_c}{\rho}}_{=:\sigma}\bigg)
\eeq
in terms of  the solution $\Phi_n(\sigma)$ (normalized in $\sigma$-space) of the auxiliary Schr\"odinger equation
\beq
\left[-\frac{\hbar^2}{2m}\frac{\partial^2}{\partial\sigma^2}+\frac{1}{2}m\omega_0^2\sigma^2+U(\sigma)\right]\Phi_n=\lambda_n\Phi_n.
\label{last}
\eeq
The strategy of invariant-based inverse engineering here is to design   $\rho$ and $q_c$ first so that $I$ and $H$ commute at initial and final times, except for launching or stopping atoms as in \citet{transport}. Then $H$ is deduced from Eq. (\ref{Vinv}). Applications will be discussed in Secs.
\ref{secEXP} and \ref{secTRA}.    
\subsection{Counterdiabatic or Transitionless Tracking Approach\label{secCD}}
For the transitionless driving or counterdiabatic  approach  
as formulated by
\citet{Berry09}, and equivalently by \citet{Rice03,Rice05,Rice08}\footnote{Berry's transitionless driving method 
is equivalent to the counterdiabatic approach of 
\citet{Rice03,Rice05,Rice08}. 
In Section \ref{alternative} we shall 
see how to further exploit this scheme together with ``superadiabatic 
iterations''.},  the starting point is a time-dependent reference Hamiltonian,
\beq
H_0(t)=\sum_n | n_0(t)\rangle  E^{(0)}_n(t) \langle n_0 (t)|. 
\eeq
The approximate time-dependent adiabatic solution of the dynamics
with $H_0$ takes the form 
\beq
\label{aa}
|\psi_n^{(ad)} (t) \rangle =  e^{i \xi_n (t)} |n_0(t)\rangle,
\eeq
where the adiabatic phase reads 
%
\beq
\xi_n (t)=-\frac{1}{\hbar} \int^t_0 dt' E^{(0)}_n(t') +  i\int^t_0 dt' \langle n_0(t')| \partial_{t'} n_0(t') \rangle.
\eeq
The approximate adiabatic vectors in Eq. (\ref{aa}) are defined differently from the dynamical modes of the previous subsection, but they may potentially coincide, 
as we shall see. 
Defining now the unitary operator
\beq
U= \sum_n e^{i \xi_n (t)} |n_0(t)\rangle \langle n_0(0)|,
\eeq
a Hamiltonian $H(t)=i \hbar \dot{U}{U}^{\dag}$ can be constructed to drive the system exactly along the adiabatic paths of $H_0(t)$, as
$H(t)=  H_{0}(t) + H_{cd}(t)$, where 
\beqa
H_{cd}(t)= i \hbar \sum_n  \bigg(|\partial_t n_0(t)  \rangle \langle n_0(t) |
-\langle n_0(t) | \partial_t n_0(t)  \rangle | n_0(t) \rangle \langle n_0(t) |\bigg)
\eeqa
is purely non-diagonal in the
$\{|n_0(t)\ra\}$ basis.

We may change the $E^{(0)}_n(t)$, 
and therefore $H_0(t)$ itself, keeping the same $|n_0(t)\ra$. We could for example make all the $E^{(0)}_n(t)$ zero, or set $\xi_n(t)=0$ \citep{Berry09}. 
Taking into account this freedom the  Hamiltonian for transitionless driving 
can be generally written as
\beqa
\label{NBH}
H (t)= -\hbar \sum_n |n_0 (t) \ra \dot{\xi}_n  \la n_0 (t)|+ i \hbar \sum_n  | \partial_t n_0(t)  \rangle \langle n_0(t)|.
\eeqa
%
%
%
Subtracting $H_{cd}(t)$, the generic $H_0$ is 
\beq
\label{general H_0}
H_0 (t)=\sum_n |n_0 (t) \ra\big[i\hbar \la n_0(t)|\partial_t n_0 (t) \ra
- \hbar \dot{\xi}_n \big]\la n_0 (t)|.
\eeq
It is usually required that $H_{cd}(t)$ vanish for $t<0$ and $t>t_f$, either suddenly or continuously at the boundary times. In that case the $\{|n_0 (t) \ra\}$ become also at the extreme times (at least at $t=0^{-}$ and $t=t_f^{+}$) eigenstates of the full Hamiltonian.

Using Eq. (\ref{invad}) and the orthonormality of the $\{|n_0(0)\ra\}$
we may write invariants of $H(t)$ with the form
$
I(t)=\sum_n |n_0(t)\ra \lambda_n\la n_0(t)|.
$
For the simple choice $\lambda_n=E^{(0)}_n(0)$, then $I(0)=H_0(0)$.

{In this part and also in Sec. \ref{inveng}} the invariant-based and
transitionless-tracking-algorithm approaches have been presented in a common language to make their relations obvious.
Reinterpreting the phases of Berry's method as $\xi_n (t)=\alpha_n (t)$, and the states as $|n_0(t)\ra=|\phi_n(t)\ra$, 
the Hamiltonians $H(t)$ in Eqs. (\ref{inHa}) and (\ref{NBH}) may be equated.
As well, the $H_0(t)$ implicit in the invariant-based method is given by Eq. (\ref{general H_0}), so that the dynamical modes can be also understood as approximate adiabatic modes of $H_0(t)$ \citep{Chen11}.
An important caveat is that the two methods could coincide but they do not have to. Given $H(0)$ and $H(t_f)$, there is much freedom to interpolate them using different invariants, phase functions, and reference Hamiltonians $H_0(t)$. In other words, these methods do not provide a unique shortcut but  families of them. This flexibility enables us to optimize the path according to physical criteria and/or  operational constraints.

{\it Non-Hermitian Hamiltonians.--}
A generalization is possible for non-Hermitian Hamiltonians in a weak non-hermiticity regime \citep{NHSara,erratum}. It was applied to engineer a shortcut laser interaction and accelerate the decay 
of a two-level atom with spontaneous decay. Note that the concept of ``population'' is problematic for non-Hermitian Hamiltonians
\citep{Jolicard}.
This affects in particular the definition of ``adiabaticity'' and 
of the shortcut concept. It is useful to rely instead on normalization-independent quantities, such as the norm of a wave-function component in a biorthogonal basis \citep{erratum}.   

{\it Many-body Systems.--}
Following \citet{DRZ12}, the transitionless quantum driving  can be extended as well to many-body quantum critical systems, exploiting recent advances in the simulation of coherent $k$-body interactions \citep{kbody,Barreiro11}. In this context 
STA allow  a finite-rate crossing of a second order quantum phase transition without creating excitations.
Consider the family of quasi-free fermion Hamiltonians in dimension $D$,
%
$\mathcal{H}_0 =\sum_{\k}\psi_{\k}^\dagger \left[ \vec{a}_\k (\lambda(t))  \cdot \vec{ \sigma}_\k \right] \psi_{\k}
$, 
%
where the $\k$-mode Pauli matrices are $\vec \sigma_{\k} \equiv (\sigma_\k^x, \sigma_\k^y,\sigma_\k^z )$  and  $\psi_{\k}^\dagger = (c_{\k,1}^\dagger,c_{\k,2}^\dagger)$ are fermionic operators, and the sum goes over independent ${\bf k}$-modes. Particular instances of quantum critical models  within this family of Hamiltonians are the Ising and XY models in $D=1$ \citep{Sachdev}, and the Kitaev model in $D =2$ \citep{EK} and  $D=1$ \citep{1dKitaev}. The function $\vec a_\k (\lambda) \equiv (a^x_\k (\lambda),a^y_\k (\lambda),a^z_\k (\lambda))$ is specific for each  model \citep{Dziarmaga10}.  All these models can be written down as a sum of independent Landau-Zener crossings, where the 
instantaneous $\k$-mode eigenstates have eigenenergies 
$
\varepsilon_{\k,\pm}=\pm |\vec a_\k(\lambda)| = \pm \sqrt{a_\k^x(\lambda)^2+a_\k^y(\lambda)^2+a_\k^z(\lambda)^2}. \nonumber
$
It is possible to adiabatically cross the quantum critical point  driving the dynamics along the  instantaneous eigenmodes of $\mathcal{H}_0$ provided that the dynamics is driven by 
the modified Hamiltonian $\mathcal{H}=\mathcal{H}_0+\mathcal{H}_{cd}$, where \citep{DRZ12}
\beqa
\mathcal{H}_{cd} &=& \lambda'(t)\sum_{\k}\frac{1}{2  |\vec a_\k(\lambda)|^2} \psi_{\k}^{\dag} \left[ (\vec a_\k (\lambda) \times \partial_\lambda  \vec a_\k (\lambda)) \cdot \vec \sigma_{\k}  \right] \psi_{\k}
\eeqa
is typically highly non-local in real spaces and involves many-body interactions in the spin representation. 
However, it was shown in the 1D quantum Ising model that a truncation of $\mathcal{H}_{cd}$ with interactions restricted to range $M$ is efficient to suppress excitations on modes $k>M^{-1}$ \citep{DRZ12}.
\subsection{Fast-forward Approach\label{secmethod}}
Based on some earlier results \citep{MNPra}, the fast-forward (FF) formalism for adiabatic dynamics and application examples were worked out in \citet{MNProc,MN11,Masuda2012}
for the Gross-Pitaevskii equation or the corresponding Schr\"odinger equation. 
The aim of the method is to accelerate a ``standard'' system subjected to a slow variation of external parameters by canceling a divergence due to an infinitely-large magnification factor 
with the infinitesimal slowness due to adiabaticity. A fast-forward potential is constructed which leads to the speeded-up evolution but, 
as a consequence of the different steps and functions introduced, the method is somewhat involved, which possibly hinders a broader application.  
The streamlined construction of 
fast-forward potentials presented in \citet{ErikFF} is followed here.  

%
The starting point is the 3D time-dependent Gross-Pitaevskii (GP) equation
\citep{Dalfovo}
\beq
\label{start}
i\hbar\frac{\partial \psi(\mathbf{x},t)}{\partial t}
= -\frac{\hbar^2}{2m} \nabla^2 \psi(\mathbf{x},t) + V(\bold{x},t)\psi(\mathbf{x},t) 
+ g_{\rm 3}|\psi(\mathbf{x},t)|^2\psi(\mathbf{x},t).
\eeq
Using the ansatz $\psi(\mathbf{x},t)=r(\bold{x},t)e^{i\phi(\mathbf{x},t)}$
($r(\mathbf{x},t), \phi(\mathbf{x},t) \in \mathbb{R}$)
we formally solve for $V(\mathbf{x},t)$ in \eqref{start} and get 
for the real and imaginary parts
\beqa
{\rm{Re}}[V(\bold{x},t)]&=&-\hbar{\dot \phi}+\frac{\hbar^2}{2m}\bigg(\frac{\nabla^2 r}{r}-(\nabla \phi)^2\bigg)-g_3r^2, \label{real}
\\
{\rm{Im}}[V(\bold{x},t)]&=&\hbar\frac{\dot r}{r}+\frac{\hbar^2}{2m}\bigg(\frac{2\nabla \phi\cdot \nabla r}{r}+\nabla^2 \phi\bigg).
\label{imag}
\eeqa
Imposing ${\rm{Im}}[V(\bold{x},t)]=0$, i.e.,  
\beq
\frac{\dot r}{r}+\frac{\hbar}{2m}\bigg(\frac{2\nabla \phi\cdot \nabla r}{r}+\nabla^2 \phi\bigg)=0,
\label{imag0}
\eeq
Eq. (\ref{real}) gives a real potential.
In the inversion protocol it is assumed that 
the full Hamiltonian and the corresponding eigenstates are known at the boundary times. Then we design $r(\bold{x},t)$,   
solve for $\phi$ in Eq. (\ref{imag0}), and finally get the potential $V$ from Eq. (\ref{real}). 
In \citet{ErikFF} it was shown how the work of \citet{MNPra,MNProc,MN11}
relates to this streamlined construction.

Since the phase $\phi$ that solves Eq. (\ref{imag0}) 
depends in general on the particular  $r({\bf{x}},t)$, Eq. (\ref{real}) 
gives in principle a 
state-dependent potential. However, in some special circumstances,
the fast-forward potential remains the same for all modes.
This happens in particular for the Schr\"odinger equation, $g_3=0$, and Lewis-Leach potentials 
associated with quadratic-in-momentum invariants. In other words, the invariant-based approach can be
formulated as a special case of the simple inverse method \citep{ErikFF}. 
\subsection{Alternative Shortcuts Through Unitary Transformations
\label{alternative}}
Shortcuts found  via the methods described so far or 
by any other approach might be difficult to implement in practice. 
In the cd approach, for instance, the 
structure of the complementary Hamiltonian $H_{cd}$ 
could be quite different 
from the structure of the reference Hamiltonian $H_0$. 
Here are three examples, the first two for a particle of mass $m$ in 1D, the third one for a two-level system: 

- Example 1: Harmonic oscillator expansions \citep{Muga10}, see Sec. \ref{secEXP}:
\beq
H_0=p^2/(2m)+m\omega^2 q^2/2,
\;\;  
H_{cd}=-(pq+qp)\dot{\omega}/(4 \omega).
\eeq
%

- Example 2: Harmonic transport with a trap of constant frequency $\omega_0/2\pi$
and displacement $q_0(t)$ \citep{transport}, see Sec. \ref{secTRA}:
\beq
H_0=p^2/(2m)+(q-q_0(t))^2m\omega_0^2/2,
\;\;  
H_{cd}=p\dot{q}_0.
\eeq
%

- Example 3: Population inversion in a two-level system \citep{Berry09,Ch10b,Sara12},
see Sec. \ref{secINT}: 
\beq
H_0=\left(
\begin{array}{cc}
Z_0&X_0
\\
X_0&-Z_0
\end{array}
\right),
\;\;  
H_{cd}=\hbar(\dot{\Theta}_0/2)\sigma_y,
\label{28}\eeq
where $\Theta_0=\arccos(Z_0/R_0)$ is a polar angle
and $R_0=(X_0^2+Z_0^2)^{1/2}$.
%

In all these examples the experimental implementation of 
$H_0$ is possible, but the realization of the 
counter-diabatic terms is problematic.
A way out is provided by unitary transformations that
generate alternative shortcut protocols without the
undesired terms in the Hamiltonian \citep{Sara12}.
{A standard tool is the use of different interaction pictures for describing one physical
setting.} Unitary operators $\mathcal{U} (t)$
connect the different pictures and the goal is frequently to 
work in a picture that facilitates the mathematical manipulations.
In this standard scenario all pictures describe the same physics, the same physical experiments and manipulations.

The main idea in \citet{Sara12} is to regard 
instead the unitary transformations as a way to generate different 
physical settings and different experiments,
not just as mathematical transformations.
%
The starting point is a shortcut described by the Schr\"odinger 
equation $i\hbar\partial_t \psi(t)=H(t)\psi(t)$, our reference protocol. 
(In all the above examples $H=H_0+H_{cd}$.)  
The new dynamics is given by  
$i\hbar\partial_t \psi'(t)=H'(t)\psi'(t)$, where $\psi'(t)={\mathcal{U}}(t)^\dagger\psi(t)$, 
and 
$H'= {\mathcal{U}}^\dagger(H-K){\mathcal{U}}$, where  $K=i\hbar\dot{{\mathcal{U}}}{\mathcal{U}}^\dagger$. 
If ${\mathcal{U}}(0)={\mathcal{U}}(t_f)=1$ the final states will coincide, 
i.e., $\psi'(t_f)=\psi(t_f)$ for a given initial state
$\psi'(0)=\psi(0)$. 
If, in addition, $\dot{{\mathcal{U}}}(0)=\dot{{\mathcal{U}}}(t_f)=0$, then 
$H(0)=H'(0)$, and $H(t_f)=H'(t_f)$.  
Let us now list the unitary transformations that provide for the three examples realizable Hamiltonians \citep{Sara12}:

- Example 1: Harmonic oscillator expansions, 
\beq
{\mathcal{U}}=\exp{\bigg(i\frac{m\dot{\omega}}{4\hbar \omega}q^2\bigg)},
\;\;  
H'=p^2/(2m)+m{\omega'}^2 q^2/2,
\label{u1}
\eeq
where
%
$
\omega'=\left[\omega^2-\frac{3\dot{\omega}^2}{4\omega^2}+
\frac{\ddot{\omega}}{2\omega}\right]^{1/2}.
$

%
- Example 2: Harmonic transport, 
\beq
{\mathcal{U}}=
\exp{(-im\dot{q}_0q/\hbar)},
\;\;
H'=p^2/(2m)+(q-q'_0(t))^2m\omega_0^2/2,
\label{u2}
\eeq
where
%
$
q_0'=q_0+\ddot{q}_0/\omega_0^2.
$
%

- Example 3: Population inversion in a two-level system,
\beq
{\mathcal{U}}=\left(
\begin{array}{cc}
e^{-i\phi/2}&0
\\
0&e^{i\phi/2}
\end{array}
\right),
\;\;  
H'=\left(
\begin{array}{cc}
Z_0-\hbar\dot{\phi}/2&P
\\
P&-Z_0+\hbar\dot{\phi}/2
\end{array}
\right),
\label{u3}
\eeq
where
%
$
\phi=\arctan{(\hbar\dot{\Theta}_0/2X_0)},\,\, 0\le\phi < 2\pi, 
$
and 
$
P=[X_0^2+(\hbar\dot{\Theta}_0/2)^2]^{1/2}.
$

Why do the ${\mathcal{U}}$s in Eqs. (\ref{u1}-\ref{u3}) have these forms?
The answer lies in the symmetry possesed by the Hamiltonian. 
Transformations of the form ${\mathcal{U}}=e^{if(t)G_j}$ based on generators $G_j$ of the corresponding Lie algebra 
produce operators within the algebra and, by suitably manipulating the
function $f(t)$ undesired terms may be eliminated.   

{\it Superadiabatic iterations.--}
As discussed in Sec. \ref{secCD},
Demirplak, Rice, and Berry proposed to add  
a suitable counterdiabatic (cd) term\footnote{This is the 
$H_{cd}$ term of Sec. \ref{secCD}. The superscript $^{(0)}$ is added now to distinguish it from  higher order cd terms introduced below.} $H_{cd}^{(0)}$ 
to the time dependent Hamiltonian $H_0(t)$ so as to follow the adiabatic dynamics of $H_0$. 
The same $H_{cd}^{(0)}$ also appears naturally when studying the adiabatic approximation of the original system, i.e., the one evolving with $H_0$.
This system behaves adiabatically, following the eigenstates of 
$H_0$, precisely when the counterdiabatic term is negligible. 

This is evident in an interaction picture (IP) based on the 
unitary transformation $A_0(t)=\sum_n |n_0(t)\ra\la n_0(0)|$
such that $\psi_1(t)=A_0^\dagger\psi_0$. 
In this IP, the new Hamiltonian is 
$H_1(t)=A_0^\dagger(t)(H_0(t)-K_0(t))A_0(t)$
and $K_0(t)=i\hbar \dot{A}_0(t)A_0^\dagger(t)$. 
If $K_0(t)$ is zero or negligible, $H_1(t)$ becomes diagonal in the basis 
$\{|n_0(0)\ra\}$, so that the IP equation is an uncoupled system
with solutions 
\beq
|\psi_1(t)\ra=\sum_n |n_0(0)\ra e^{-\frac{i}{\hbar}\int_0^t E_n^{(0)}(t')dt'}
\la n_0(0)|\psi_1(0)\ra.
\eeq
Correspondingly, 
%
$
|\psi_0(t)\ra=\sum_n |n_0(t)\ra e^{-\frac{i}{\hbar}\int_0^t E_n^{(0)}(t')dt'}
\la n_0(0)|\psi_0(0)\ra.
$
%

The same solution, which, for a non-zero $K_0$, is only approximate, is found exactly by adding to the IP Hamiltonian 
the counterdiabatic term $A_0^\dagger(t)K_0(t)A_0(t)$. 
This requires an external intervention and changes the physics of the  original system. In the IP the modified Hamiltonian is  $H^{(1)}\equiv H_1+A_0^\dagger(t)K_0(t)A_0(t)=
A_0^\dagger(t)H_0(t)A_0(t)$ and in  
the Schr\"odinger picture (SP)
the modified Hamiltonian is $H_0^{(1)}(t)=H_0(t)+K_0(t)$, so we identify $H_{cd}^{(0)}(t)=K_0(t)$.
In other words, a ``small'' coupling term $K_0$ that makes the adiabatic approximation a good one also implies a small 
counterdiabatic manipulation. However, irrespective of the size of $K_0$,  $H_0^{(1)}(t)$ provides a shortcut to slow adiabatic following because 
it keeps the populations in the instantaneous basis of $H_0$ invariant, 
in particular at the final time $t_f$.  

Looking for generalized adiabatic approximations, 
\citet{Garrido}, \citet{Berrysa} or \citet{NMR} have investigated further iterative interaction pictures
and the corresponding approximations. The idea is best understood by working 
out explicitly the next iteration: one starts with 
$i\hbar\partial_t \psi_1(t)=H_1\psi_1(t)$ and diagonalizes $H_1(t)$
to produce its eigenbasis $\{|n_1(t)\ra\}$.
A unitary operator $A_1=\sum_n|n_1(t)\ra\la n_1(0)|$ plays now the same 
role as $A_0$ in the previous IP. It defines a new IP wave function $\psi_2(t)=A_1^\dagger(t)\psi_1$ that satisfies 
$i\hbar\partial_t \psi_2(t)=H_2\psi_2(t)$, where $H_2(t)=A_1^\dagger(t)(H_1(t)-K_1(t))A_1(t)$ and $K_1=i\hbar\dot{A}_1A_1^\dagger$. 
If $K_1$ is zero or ``small'' enough, i.e. if a (first order) 
superadiabatic approximation 
is valid, 
the dynamics would be uncoupled in the new interaction picture, namely, 
\beq\label{psi2unc}
|\psi_2(t)\ra=\sum_n |n_1(0)\ra e^{-\frac{i}{\hbar}\int_0^t E_n^{(1)}(t')dt'}
\la n_1(0)|\psi_2(0)\ra.
\eeq
We may get the same result by changing the physics and adding  
$A_1^\dagger(t)K_1(t)A_1(t)$ to $H_2$ \citep{Rice08,Sara12}.
In the SP 
the added interaction becomes a first order counterdiabatic term $H_{cd}^{(1)}=A_0 K_1 A_0^\dagger$.  
Transforming back to the  
SP and using $A_j(0)=1$ the state (\ref{psi2unc}) becomes 
\beq
|\psi_0(t)\ra=\sum_n\sum_m |m_0(t)\ra \la m_0(0)|n_1(t)\ra e^{-\frac{i}{\hbar}\int_0^t E_n^{(1)}(t')dt'}
\la n_1(0)|\psi_0(0)\ra.
\eeq
%
Quite generally the populations of the final state in the adiabatic basis $\{|n_0(t_f)\ra\}$ will be different 
from the ones of the adiabatic process, unless $|n_0(0)\ra=|n_1(t_f)\ra$
and $|n_1(0)\ra=|n_0(0)\ra$, up to phase factors. 
The first condition is satisfied if $K_0(t_f)=0$ and the second one if 
$K_0(0)=0$. 
Then the superadiabatic process will actually lead to the same final populations
as an adiabatic one, possibly with different phases for the 
individual components. Similarly, the first-order counterdiabatic term 
$H_{cd}^{(1)}$ would provide a shortcut with $H_0^{(2)}=H_0+H_{cd}^{(1)}$ in the SP,
different from the one carried out by $H_{0}^{(1)}$.  
Moreover, if $K_1(0)=K_1(t_f)=0$, then $H_0^{(2)}=H_0$, at $t=0,t_f$.    
Further iterations
define higher order  superadiabatic frames. 
Is there any advantage in using one or another counterdiabatic scheme?
There are two reasons that could make higher order schemes attractive
in practice: one is that the structure of the $H_{cd}^{(j)}$ may change with $j$. For example, for a two-level population inversion 
$H_{cd}^{(0)}=\hbar(\dot{\Theta}_0/2)\sigma_y$, whereas 
$H_{cd}^{(1)}=\hbar(\dot{\Theta}_1/2)(\cos\Theta_0 \sigma_x-\sin\Theta_0\sigma_z)$, where $\Theta_1$ is the polar angle 
corresponding to the the Cartesian components of
$H_1=X_1\sigma_x+Y_1\sigma_y+Z_1\sigma_z$ \citep{Sara12}.  
The second reason is that, for a fixed process time, the cd-terms are 
smaller in norm as $j$ increases, up to a value
in which they begin to grow, see e.g. \citet{NMR}.
One should pay attention though not only to the size of the cd-terms but also to the feasibility of the boundary conditions at the time edges to really generate shortcuts in this manner. 
\subsection{Optimal Control Theory}
Optimal control theory (OCT) is a vast field covering many techniques and applications. As for STA, fast expansions \citep{Salamon09}, wavepacket splitting \citep{S07,S09a,S09b},  transport \citep{Calarco} and many-body state preparation \citep{RC09} have been addressed 
with different OCT approaches. The  
combination of OCT techniques with invariant-based engineering STA is 
particularly fruitful since the later provides by construction families of 
protocols that achieve a perfect fidelity or vanishing final excitation, 
whereas OCT may help to select among the many possible protocols the ones
that optimize some physically relevant variable \citep{Li10,Li11,OCTtrans,Li12}. 
In this context the theory used so far is the maximum principle of \citet{LSP}. 
For a dynamical system 
$\dot{x}=\mathbf{f}(\mathbf{x}(t),u)$, where $\mathbf{x}$ is the state vector and $u$ the scalar control, {in order} to minimize the cost function $J(u)=\int_0^{t_f}
g(\mathbf{x}(t),u)dt$, the principle states that the coordinates of the extremal vector 
$\mathbf{x}(t)$ and of the corresponding adjoint state $\mathbf{p}(t)$ formed by Lagrange multipliers, fulfill Hamilton's equations for a control Hamiltonian $H_c=p_0g(\mathbf{x}(t),u)+\mathbf{p}^T\cdot \mathbf{f}(\mathbf{x}(t),u)$. For almost all times during the process
$H_c$ attains its maximum at $u=u(t)$ and $H_c=c$, where $c$ is constant.    
We shall discuss specific applications in Secs. \ref{secEXP} and \ref{secTRA}.
%
%
%
%
%
%
\section{Expansions of trapped particles\label{secEXP}}
Performing fast expansions of trapped cold atoms without losing or exciting them is important for many applications: 
for example to  reduce velocity dispersion and collisional shifts in  spectroscopy and atomic clocks,
decrease the temperature,
adjust the density to avoid three body losses,
facilitate temperature and density measurements,
or to change the size of the cloud for further manipulations.
Of course trap compressions are also quite common. 

For harmonic traps  
we may address expansion or compression processes with the 
quadratic-in-$p$ invariants theory by setting $q_c=U=F=0$ in Eq. (\ref{Vinv}).
This means that Eq. (\ref{alphaeq}) does not play any role and the important auxiliary equation is the ``Ermakov equation'' (\ref{Erma}). The physical meaning of $\rho$ is
determined by its proportionality to the standard deviation of the position 
of the ``expanding
(or contracting) modes'' $e^{i\alpha_n}\phi_n$.

Here we shall discuss the expansion from $\omega(0)=\omega_0$ to $\omega(t_f)=\omega_f$ \citep{Ch10}.
%
Choosing 
\beq
\label{bct0}
\rho(0)=1,\;\; \dot{\rho}(0)=0,
\eeq
$H(0)$ and $I(0)$ commute. They actually become equal, and have common eigenfunctions.
Consistent with the Ermakov equation, $\ddot{\rho}(0)=0$ holds as well for a continuous frequency. 
At $t_f$ we impose\footnote{If $\ddot{\rho}(t_f)\ne 0$ the final frequency would not be $\omega_f$ but $\omega(t_f)=[\omega_f^2-\ddot{\rho}/\gamma]^{1/2}$. If discontinuities are 
allowed and the frequency is changed abruptly from $\omega(t_f)$ to $\omega_f$
the excitations will also be avoided, at least in principle.
A similar discontinuity is possible at $t=0$ if $\ddot{\rho}(0)\ne 0$ and 
the frequency jumps abruptly from $\omega_0$ to $\omega(0)=[\omega_0^2-\ddot{\rho}(0)]^{1/2}$.}  
\beq
\rho(t_f)=\gamma=(\omega_0/\omega_f)^{1/2},\;\;
\dot{\rho}(t_f)=0,\;\; \ddot{\rho}(t_f)=0.
\label{bctf}
\eeq
In this manner the expanding mode is an instantaneous eigenvector of $H$ at $t=0$ and $t_f$, 
regardless of the exact form of $\rho(t)$.
To fix $\rho(t)$, one chooses a functional form to interpolate
between these two times, flexible enough to satisfy the boundary conditions. 
For a simple polynomial ansatz 
%
%
%
$
\rho (t) =
6 \left(\gamma -1\right) s^5
-15 \left(\gamma-1\right) s^4 +10 \left(\gamma-1\right)s^3
+ 1
$ \citep{Palao}, 
%
where  $s=t/t_f$.

The next step is to solve for $\omega(t)$
in Eq. (\ref{Erma}).
This procedure poses no fundamental lower limit to $t_f$, which could be in principle arbitrarily small. There are nevertheless practical limitations 
and/or prices to pay. For short enough $t_f$, $\omega(t)$ may become purely imaginary
at some $t$ \citep{Ch10} and the potential becomes a parabolic repeller. Another difficulty is that the transient energy required may be too high, as 
discussed in \citet{energy} and in the following subsection.
Since actual traps are only approximately harmonic,  large transient energies will imply perturbing effects of anharmonicities and thus undesired
excitations of the final state, or even atom losses.
\subsection{Transient Energy Excitation}
Knowing the transient excitation energy is also important to quantify the principle of unattainability of zero temperature, first
enunciated by Nernst.
This principle is usually formulated as the impossibility to reduce the temperature of any system to the absolute zero in a finite number of operations,
and identified with the third law of thermodynamics.  
Kosloff and coworkers  in \citep{Salamon09} have restated the unattainability principle in quantum refrigerators as the vanishing
of the cooling rate when the temperature of the cold bath approaches zero, and quantify it by the scaling law that relates
cooling rate and cold bath temperature.
We shall examine here the consequences of the transient energy excitation
on the unattainability principle in two ways: for a single, isolated expansion, and considering  the expansion as one of the branches of a quantum refrigerator cycle \citep{energy}.

A lower bound ${\mathcal B}_n$ for the time-averaged energy of the n-th expanding mode $\overline{E_n}$ 
(time averages from $0$ to $t_f$ will be denoted by a bar) is found by applying calculus of variations \citep{energy},
so that $\overline{E_n}\ge{\mathcal B}_n$. 
If the final frequency $\omega_f$ is small enough to satisfy $t_f \ll 1/\sqrt{\omega_0 \omega_f}$, and $\gamma \gg 1$, the
lower bound has the asymptotic form
$
{\mathcal B}_n\approx {(2n+1) \hbar}/{(2 \omega_f t^2_f)}.
$
A consequence is that 
$
t_f \geq \sqrt{(2n+1) \hbar/(2 \omega_f \overline{E_n})}.
$
When $\overline{E_n}$ is limited, because of anharmonicities or a finite
trap depth, the scaling is fundamentally the same as the one found for  bang-bang methods with real frequencies \citep{Salamon09}, and leads to a cooling rate $R\propto T_c^{3/2}$ in an inverse  quantum Otto cycle (the proportionality factor may be improved by increasing
the allowed $\overline{E_n}$). This dependence had been previously conjectured to be a universal one  characterizing the unattainability principle for any cooling cycle \citep{KoEPL09}. The results in \citet{energy} provide strong support for the validity of this conjecture
within the set of processes defined by ordinary harmonic 
oscillators with time-dependent frequencies. In \citep{EPL11} a faster rate $\sim -T_c/\log T_c$ is found 
with optimal control techniques for bounded trap frequencies, 
allowed to become imaginary. There  is no contradiction with the 
previous scaling since 
bounding the trap frequencies does not bound the system energy. 
In other words, achieving such fast cooling is not possible if the  
energy cannot become arbitrarily large.     

Independently of the participation of the harmonic trap expansion as a
branch in a refrigerator cycle, we may apply the previous analysis also to a single expansion, assuming that the initial and final states are  canonical density operators characterized by temperatures $T_0$ and $T_f$. These are 
related by $T_f=(\omega_f/\omega_0)T_0$ for a population-preserving process.
{}In a
harmonic potential expansion, the unattainability of a zero
temperature can be thus reformulated
as follows:
The transient excitation energy becomes infinite for any
population-preserving and finite-time process when the final temperature is zero (which requires $\omega_f=0$). The excitation energy has to be provided by an external device, so a fundamental obstruction to reach $T_f=0$ in a finite time, is the need for a source of infinite power \citep{energy}.


The standard deviation of the energy was also studied numerically \citep{energy}. { There it was found that the dominant dependences of the time averages scale with} $\omega_f$ and $t_f$ in the same way as the average energy. These dependences 
are different from the ones in the \citet{AA} (AA) relation 
$
\overline{\Delta H}\,   t_f \geq  \frac{h}{4},
$
where
$
\overline{{\Delta H}}= {\int^{t_f}_0 \Delta H (t) dt}/{t_f}.
$ 
%
%
%
\subsection{Three Dimensional Effects}
The previous discussion is limited to one dimension (1D) but actual traps are 
three-dimensional and at most effectively 1D.
\citet{3d} worked out the theory and 
performed numerical simulations of     
fast expansions of cold atoms in a three-dimensional Gaussian-beam optical trap. Three different methods to avoid final motional excitation were compared: 
inverse engineering using Lewis-Riesenfeld invariants, which provides the best overall performance, a bang-bang approach with one intermediate frequency, and a ``fast adiabatic approach''\footnote{
The adiabaticity condition for the harmonic oscillator is  
$|\sqrt{2}\dot{\omega}/(8 \omega^2)|\ll 1$. An efficient, but still adiabatic, strategy by \citet{Ch10} is to distribute $\dot{\omega}/\omega^2$ uniformly along the trajectory, i.e., $\dot{\omega}/\omega^2=c$, $c$ being constant. 
Solving this differential equation and imposing $\omega_f=\omega(t_f)$ we get $\omega(t)=\omega_0/[1-(\omega_f-\omega_0)t/(t_f\omega_f)]$. This 
may be enough for some applications. 
This function was successfully applied in \citet{Bowler}.}.

The optical trap considered in \citet{3d} is formed by a laser, red detuned with respect to an atomic transition, and is characterized in the harmonic approximation by longitudinal and radial frequencies. To fourth order in the coordinates the effective potential includes anharmonic terms and radial-longitudinal coupling terms.   
While magnetic traps allow for an independent control of longitudinal and radial frequencies \citep{Nice10,Nice11,Nice11b}, this is not the case for a simple laser trap.
In \citet{3d} it was assumed that the time-dependence of the longitudinal frequency is engineered to avoid final excitations with a simple 1D harmonic theory.
The main conclusion of the study is that the transitionless expansions 
in optical traps are feasible under realistic conditions.
For the inverse engineering method, the main perturbation is due to the 
possible adiabaticity failure in the radial direction, which can be suppressed or 
mitigated by increasing the laser waist. This waist increase would also reduce 
smaller perturbing effects due to longitudinal anharmonicity or
radial-longitudinal coupling. 
The simple bang-bang approach fails because the time for the radial expansion is badly mismatched with respect to the ideal time, and   
the fast adiabatic method fails for short expansion times as a result of 
longitudinal excitations.  
Complications such as perturbations due to different noise types, and consideration of condensates, gravity effects, or 
the transient realization of imaginary trap frequencies
are still open questions. 
Other extensions of \citet{3d} could involve the addition of a second laser for further control of the potential shape, or alternative trap shapes.
Optical traps based on Bessel laser beams, for example, may be useful 
to decouple longitudinal and radial motions. 
\subsection{Bose-Einstein Condensates\label{econd}}
In this section we shall discuss the possibility of realizing STA in a harmonically trapped Bose-Einstein condensate
using a scaling ansatz.
A mean-field description of this state of matter is based on the time-dependent Gross-Pitaevskii equation (GPE) \citep{Dalfovo}, 
%
\beqa
i\hbar\frac{\partial\Psi(\x,t)}{\partial t}=\bigg[-\frac{\hbar^2}{2m}\Delta+\frac{1}{2}m\om^2(t)\x^2+g_D|\Psi(\x,t)|^2\bigg]\Psi(\x,t).
\eeqa
Here, $\Delta$ is the $D$-dimensional Laplacian operator and $g_D$ is the $D$-dimensional  coupling constant. 
For a three-dimensional cloud, using the normalization $\int|\Psi(\x,t)|^2d\x=1$, $g_{\rm 3}=\frac{4\pi\hbar^2Na}{m}$, for a condensate of 
a number of atoms $N$ of mass $m$, interacting with each other  through a contact Fermi-Huang pseudopotential parameterized by a $s$-wave scattering length $a$. 
In $D=1,2$ the corresponding expression for $g_{D}$ can be obtained by  a dimensional reduction of the 3D  GPE
\citep{Salas}.
As a mean-field theory the GPE
overestimates the phase coherence of real Bose-Einstein condensates. 
The presence of phase fluctuations generally induces a breakdown of the dynamical  self-similar scaling law that governs the dynamics of the expanding cloud 
and the formation of density ripples. 
The conditions for quantum phase fluctuations to be negligible 
for STA were discussed in \citet{delcampo11a}.
In the following we shall ignore phase fluctuations. The results of this section will be generalized to strongly correlated gases in Sec.
\ref{mbsta}, including as a particular case, the microscopic model of ultracold bosons interacting through s-wave scattering.

STA in the mean-field 
regime were designed in \citet{MCRG09} based on the classic results by \citet{CD96}, and  \citet{KSS96},
who found the exact dynamics of the condensate wavefunction under a time-modulation of the harmonic trap frequency. 
Consider a condensate wavefunction $\Psi(\x,t=0)$, a solution of the time-independent GPE 
with chemical potential $\mu$ in a harmonic trap of frequency
$\om_0$, i.e., $(-\frac{\hbar^2}{2m}\Delta+\frac{1}{2}m\om_0^2\x^2+g_D|\Psi(\x,t=0)|^2-\mu)\Psi(\x,t=0)=0$.
Under a modulation of the trap frequency $\om(t)$ the scaling ansatz
\beqa
\label{tdbec}
\Psi(\x,t)=\frac{1}{\rho^{\frac{D}{2}}}\exp\bigg[i\frac{m|\x|^2}{2\hbar}\frac{\dot{\rho}}{\rho}-i\frac{\mu\tau(t)}{\hbar}\bigg]\Psi\left(\frac{\x}{\rho},t=0\right)
\eeqa
is an exact solution of the time-dependent Gross-Pitaevskii equation provided that 
\beqa
\ddot{\rho}+\om(t)^2 \rho=\frac{\om_0^2}{\rho^3}, \qquad g_D(t)=\frac{g_D(t=0)}{\rho^{2-D}},\qquad \tau(t)=\int_0^t \frac{dt'}{\rho^2}.
\eeqa
It follows that the scaling factor $\rho$ must be a solution of the  Ermakov equation, precisely as in the single-particle harmonic oscillator case.
This paves the way to engineer a shortcut to an adiabatic expansion or compression from the initial state $\Psi(\x,t=0)$ to a target state 
$\Psi(\x,t_f)=\Psi(\x/\rho,t=0)/\rho^{\frac{D}{2}}$ by designing the trajectory $\rho(t)$.
The modulation of the coupling constant required in $D=1,3$ can be implemented with the aid of a Feshbach  resonance \citep{MCRG09}, 
or,  in $D=1$, by a modulation of the transverse confinement \citep{Staliunas04,Engels07,delcampo11a}.
The $D=2$ requires no tuning in time of the coupling constant as a result of the Pitaevskii-Rosch symmetry \citep{PR}. 
It has recently been suggested that this symmetry is broken upon quantization, constituting an instance of a quantum anomaly in ultracold gases \citep{OPL10}. 
To date no experiment has provided evidence in favor of this observation. We point out that observing a breakdown of shortcuts to expansions of 2D BEC clouds 
would help to verify this quantum-mechanical symmetry breaking.

An important simplification occurs in the Thomas-Fermi regime, where the mean-field energy  dominates over the kinetic part.
Assuming the validity of this regime along the dynamics, the scaling ansatz   (\ref{tdbec}) becomes exact as long as the following consistency equations are satisfied,
\beqa
\ddot{\rho}+\om(t)^2 \rho=\frac{\om_0^2}{\rho^{D+1}}, \qquad g_D(t)=g_D(t=0),\qquad \tau(t)=\int_0^t \frac{dt'}{\rho^D}.
\eeqa
Hence, in the Thomas-Fermi regime, it is possible to engineer a shortcut  exactly, while keeping the coupling strength $g_D$  constant \citep{MCRG09}. Optimal control theory has been recently applied in this regime to find optimal protocols 
with a restriction on the allowed frequencies \cite{Li12}. 

{\it Dimensional reduction and modulation of the non-linear interactions.--}
For low dimensional BECs, tightly confined in one or two directions,
an effective tuning of the coupling constant can be achieved by
modulating the trapping potential along the tightly confined axis, see e.g.
\citet{Staliunas04}, a proposal experimentally explored in \citet{Engels07}. 
In a nutshell, the tightly confined degrees of freedom decoupled from the weakly confined ones,
are governed to a good approximation by a non-interacting Hamiltonian.
It is then possible to perform a dimensional reduction of the 3D GPE,
and derive a lower-dimensional version for the weakly confined degrees of freedom, where
the effective coupling constant  inherits  a dependence of the width of the transverse modes which have been integrated out.
Adiabatically tuning the transverse confinement leads to a controlled tuning of the effective coupling constant.
A faster-than-adiabatic modulation can be engineered by implementing a shortcut  in the transverse degree of freedom.
Consider the 3D mean-field description
\beqa
\label{ad_3DGPE}
i\hbar\frac{\partial\Psi(\x,t)}{\partial t}=\big[-\frac{\hbar^2}{2m}\Delta+{\rm V^{ex}}(\x,t) +g_{\rm 3}|\Psi(\x,t)|^2\big]\Psi(\x,t),
\eeqa
with
$
{\rm V^{ex}}(\x,t)=\frac{m}{2}[\om_x(t)^2x^2+\om_y(t)^2y^2+\om_z(t)^2z^2]
$.
%
For tight transverse confinement ($\om_x\sim\om_y\gg \om_z$ and $N|a|\sqrt{m\om_z/\hbar}\ll 1$),  
the transverse excitations are frozen.
The transverse mode can be approximated by the single-particle harmonic oscillator ground state $\Phi_0(x,y,t)$, so that 
the wavefunction factorizes $\Psi(\x,t) = \Phi_0(x,y,t)\psi(z,t)$. 
Integrating out the transverse modes, and up to a time-dependent constant which can be gauged away, one obtains the reduced GPE 
\beqa
i\hbar\frac{\partial\psi(z,t)}{\partial t}=
\big[-\frac{\hbar^2}{2m}\frac{\partial^2}{\partial z^2}
+{\rm V^{ex}}(z)+g_{\rm 1}(t)|\psi(z,t)|^2\big]\psi(z,t),
\eeqa 
with the effective coupling
%
$g_{\rm 1}(t)
=g_{\rm 3}\iint\!\! dx dy |\Phi_0(x,y,t)|^4.$
%
A general trajectory $g_{1}(t)$ can be implemented by modifying the frequency $\om_{\perp}(t)$ of the transverse confinement according to 
\beqa
\om_{\perp}^2(t)=\om_{\perp}^2(0)\Bigg[\frac{g_{1}(t)}{g_{1}(0)}\Bigg]^2+\frac{1}{2}\frac{\ddot{g}_{1}(t)}{g_{1}(t)}-\frac{3}{4}\Bigg[\frac{\dot{g}_{1}(t)}{g_{1}(t)}\Bigg]^2
\eeqa
in quasi-1D atomic clouds~\citep{delcampo11a}. 
The first term in the RHS corresponds to the adiabatic tuning discussed in \citet{Staliunas04,Engels07} while the remaining terms are associated with the STA dynamics in the transverse modes. 
A similar analysis applies to the control of the effective coupling constant in  a pancake condensate, in the x-y plane, under tight confinement along the $z$-direction \citep{delcampo11a}.

We note that  this technique is restricted to tune the amplitude of the coupling constant, at variance with alternative techniques based on Feschbach or confinement-induced resonances which can change both the amplitude and character of the interactions, e.g.,  from attractive to repulsive \citep{BDZ08}.
\subsection{Strongly Correlated Gases}\label{mbsta}
The preceding sections were focused on single-particle systems and a mean-field description of Bose-Einstein condensates.
We have seen that the inversion of scaling laws is a powerful technique to design STA in those processes where the dynamics is self-similar, e.g. expansions, or transport.
In the following we focus on the engineering of STA in strongly correlated quantum fluids
of relevance to ultracold gases experiments.
We shall consider a fairly general model in dimension $D$ consisting of $N$ indistinguishable particles with coordinates $\x_i\in\mathbb{R}^D$,
trapped in a time-dependent isotropic harmonic potential of frequency $\om(t)$ and interacting with each other through a two-body potential ${\rm V}(\x_i-\x_j)$. 
The many-body Hamiltonian describing this system reads \citep{delcampo11}
\beqa
\label{ad_Hamiltonian}
\mathcal{H}\!=\!\sum_{i=1}^{N}\!\bigg[\!-\frac{\hbar^2}{2m}\Delta_{i}+\frac{1}{2}m\om^2(t)\x_i^2\bigg]\!+\!\epsilon\sum_{i<j}{\rm V}(\x_i-\x_j),
\eeqa
where  $\Delta_{i}$ is the $D$-dimensional Laplacian operator for the $\x_i$ variable, and $\epsilon=\epsilon(t)$ is a dimensionless time-dependent coupling strength satisfying $\epsilon(0)=1$.
We shall further assume that 
$
{\rm V}(\lambda\x)=\lambda^{-\alpha}{\rm V}(\x)
$
under scaling of the coordinates. 
Specific realizations of this model include  the Calogero-Sutherland model \citep{Sutherland98}, the Tonks-Girardeau gas \citep{OS02,MG05},  Lieb-Liniger gas \citep{BPG08}, Bose-Einstein condensates (BEC) \citep{CD96,KSS96,MCRG09}, including dipolar interactions \citep{dipolar}, and more general many-body quantum systems \citep{GBD10}.
For simplicity, we leave out other cases to which similar techniques can be applied, such as strongly interacting mixtures \citep{MG07} or systems with internal structure \citep{etonks1,etonks3}.

Let us now consider an equilibrium state $\Phi$ of the system (\ref{ad_Hamiltonian}) at $t=0$ with chemical potential $\mu$. For compactness we shall use the notation
$\x_{j:k}\equiv\{x_j,x_{j+1},\dots,x_{k-1},x_k\}$.
It is possible to find a self-similar scaling solution of the form
\beqa
\label{ad_scaling}
\Phi\left(\x_{1:N},t\right)=\frac{1}{\rho^{D/2}}\exp\bigg[i\sum_{i=1}^N\frac{ m\x_i^2\dot{\rho}}{2\rho\hbar}-i\mu\tau(t)/\hbar\bigg]
\Phi\left(\frac{\x_{1:N}}{\rho},t=0\right),
\eeqa
where $\tau(t)=\int_{0}^tdt'/\rho^2(t')$, whenever the scaling factor $\rho=\rho(t)$ is the  solution of the Ermakov differential equation, $\ddot{\rho}+\om^2(t)\rho=\om_0^2/\rho^3
$,  with $\om_0=\om(0)$, satisfying the boundary conditions $\rho(0)=1$ and $\dot{\rho}(0)=0$.
This is the same consistency equation that arises in the context of the single-particle time-dependent harmonic oscillator.

Scaling laws greatly simplify the dynamics of quantum correlations. Let us consider the time-evolution of the $n$-particle reduced density matrix
\beqa
g_n(\x_{1:n};\x_{1:n}';t)=\frac{N!}{(N-n)!}
\!\int \!
\prod_{i=n+1}^N\!d x_i \Phi^*(\x_{1:N};t)\Phi(\x_{1:n}',\!\x_{n+1:N};t).
\eeqa

Provided the scaling law holds, its time evolution reads
\beqa
g_n(\x_{1:n};\x_{1:n}';t)=
\rho^{-nD} g_n\!\left(\frac{\x_{1:n}}{\rho};\frac{\x_{1:n}'}{\rho}; 0\right)
\exp\!\left(\!-\frac{i}{\rho}\frac{\dot{\rho}}{\omega_0}
\frac{\sum_{i=1}^n(\x_i^2-\x_i'^2)}{2l^2_0}\right),
\eeqa
where $l_0=\sqrt{\hbar/m\om_0}$.

Local quantum correlations depend exclusively on the diagonal elements of $g_n(\x_{1:n};\x_{1:n}';t)$ and manifest directly
 the self-similar dynamics. For instance,  the time evolution of the density profile $n(\x)=g_1(\x;\x)$ reads
$
n(\x,t)=\rho^{-nD}n\left(\frac{\x}{\rho},t=0\right).
$

The dynamics of non-local correlations is more involved due to the presence of the oscillatory phase.
As an example,  the evolution of the one-body reduced density matrix (OBRDM) under self-similar dynamics  \citep{MG05,GBD10},
\begin{equation}
  \label{ad_g1t}
  g_1(\x,\y;t) = \frac{1}{\rho^D} g_1\left(\frac{\x}{\rho},\frac{\y}{\rho};0\right)
  \exp\left(-\frac{i}{\rho}\frac{\dot{\rho}}{\omega_0}
  \;\frac{\x^2-\y^2}{2l^2_0}\right),
\end{equation}
induces a non-self-similar evolution of the momentum distribution, its Fourier transform 
$
n(\k,t) = \int d\x d\y
\,e^{i\k\cdot(\x-\y)} g_1 (\x,\y;t).
$
 %
%
It is expected that the oscillatory phases distort quantum correlations.
The case of a free expansion, where the frequency modulation in terms of the Heaviside function $\Theta(t)$ reads 
$\om(t)=\om_0\Theta(-t)$, has received much attention.
The solution to the Ermakov equation for the scaling factor is $\rho(t)=\sqrt{1+\om_0^2t^2}$ and for
$t\gg\om_0^{-1}$, $\rho(t)\sim\om_0t$, $\dot{\rho}=\om_0$. Using the method of the stationary phase, it follows that
\beqa
n(\k,t)\sim |2\pi\om_0\l_0^2/\dot{\rho}|^Dg_1(\om_0 \k\l_0^2/\dot{\rho},\om_0 \k\l_0^2/\dot{\rho}),
\eeqa
i.e., the asymptotic momentum distribution  is mapped to the scaled density profile of the initial state \citep{Hrvoje1,Hrvoje2,GBD10}.
As a result, all information  of the off-diagonal elements of the OBRDM is lost.
Similar effects result in an expansion in finite-time $t_f\sim \om_0^{-1}$ and signal the breakdown of adiabaticity.
Excitations manifest as well in local correlation functions, e.g.  excitation of the breathing mode of the cloud.

In the adiabatic limit ($\tau\gg \om_0^{-1}$), the time-variation of the  the  scaling factor vanishes $\dot{\rho}(t)\approx 0$, resulting in the adiabatic trajectory 
$\rho(t)=\sqrt{\om_0/\om(t)}$. At all times the time-evolution of the OBRDM  and the momentum distribution can be related by a scaling transformation of their form at $t=0$, 
\begin{equation}
  \label{ad_sg1}
  g_1(\x,\y;t) = \frac{1}{\rho^D(t)} g_1\left(\frac{\x}{\rho(t)},\frac{\y}{\rho(t)};0\right),\qquad  n(\k,t) =  \rho^D(t) n(\rho(t) \k,0).
\end{equation}
These expressions can be applied for expansions ($\rho(t)>1$) and compressions ($\rho(t)<1$), and  generally still require tuning the interaction coupling strength.
Nonetheless, the required adiabatic time scale can be exceedingly long and we next tackle the problem of achieving a final scaled state in a predetermined expansion time $t_f$.
The upshot of the frictionless dynamics is that quantum correlations at the end of the quench ($t=t_f$, and only then) are those of the initial state scaled by a factor $\rho(t_f)=\gamma$ \citep{delcampo11}. In particular,
\beqa
\label{ad_stacorr}
g_1(\x,\y;t_f) = \frac{1}{\gamma^D} g_1\left(\frac{\x}{\gamma},\frac{\y}{\gamma};0\right),\qquad  n(\k,t_f) = \gamma^D n(\gamma \k,0).
\eeqa
Similar expressions hold for higher-order correlations, i.e.
$g_n(\x_{1:n},\y_{1:n};t_f) = \gamma^{nD} g_n\left(\x_{1:n}/\gamma,\y_{1:n}/\gamma;0\right)$.
Moreover, as long as the initial state is an equilibrium state in the initial trap, so it is the state at $t_f$ with respect to the final trap, preventing any non-trivial dynamics after the quench, for $t>t_f$ if $\om(t>t_f)=\om_f$.
Nonetheless, at intermediate times $t\in [0,t_f)$ the momentum distribution exhibits a rich non-equilibrium dynamics, and can show for instance, evolution towards the scaled  density profile of the initial state.

We close this section with two comments.
First, the applicability of STA based on inversion of scaling laws is not restricted to fermionic or bosonic systems, but can be applied as well to anyonic quantum fluids for which dynamical scaling laws are known \citep{delcampo08}. Systems with quantum statistics smoothly extrapolating between bosons and fermions might be realized in the laboratory following \citet{KLMR11}.
Second, the possibility of scaling up the system while preserving quantum correlations constitutes a new type of microscopy of quantum correlations in quantum fluids \citep{delcampo11,DB12}. It is as well of interest to design new protocols to reconstruct the initial quantum state of the system from the time-evolution of its density profile \citep{BB87,LS97}, a tomographic technique demonstrated experimentally in \citet{KPM97}, and applicable to many-body systems \citep{DMM08}.

{\it Scaling laws in other trapping potentials.--}
Scaling laws for many-body systems can be found for more general types of confinements.
Among them, homogeneous potentials are of particular interest, since they simplify the correspondence between ultracold atom experiments and condensed matter theory. 
The early experimental implementations of the paradigmatic particle in a box aimed at the creation of  optical billiards for ultracold gases \citep{prepainters1,prepainters2}.
Trapping of a BEC in an all-optical box was reported in \citet{becbox} and analogous traps have been created in atom chips \citep{boxchip}. 
For the purpose of implementing  STA,
the dynamical optical dipole potential may be
realized using the highly versatile ``painting technique'', which
creates a smooth and robust time-averaged potential with a
rapidly-moving laser beam \citep{painters},  or alternatively, by
spatial light modulators \citep{modu}. 

The breakdown of adiabaticity in a time-dependent homogeneous potential leads to quantum transients  related to the diffraction in time (DIT) effect, see \citet{DGCM09} for a review. 
A sharply localised matter-wave in a region of space, after sudden removal of the confinement, exhibits during free evolution density ripples.
The earliest example discussed by \citet{Moshinsky52} the free evolution of a truncated cut-off plane wave, exhibits an oscillatory pattern with the same functional form than the diffraction pattern of a classical light beam  from a semi-infinite plane. 
The phenomenon is ubiquitous in matter-wave dynamics induced by a quench, and in particular, it arises in time-dependent box potentials in  one \citep{GK76,Godoy02,DM06}, two and three \citep{Godoy03} dimensions. The effect manifests as well in strongly interacting gases such as ultracold bosons in the Tonks-Girardeau regime \citep{DM06,delcampo08}.
Moreover, when the piston walls move at a finite speed $v$, the adiabatic limit is not approached monotonically as $v\rightarrow 0$. It was shown in \citet{DMK08,Mousavi12,Mousavi12b} that an enhancement of DIT occurs 
when the walls move with the dominant velocity component of the initial confined state, due to a constructive interference between the expanding and reflected components from the walls. 
When the reflections from the confinement walls dominate, the non-adiabatic dynamics in time-dependent homogeneous potentials lead to Talbot oscillations and 
weave a quantum carpet in the time evolution of the density profile \citep{Schleich1,Schleich2}. 
Suppression of these excitations is dictated by the adiabatic theorem both in the non-interacting \citep{boxlaws1,boxlaws2,boxlaws3,bookinv} and mean-field regime \citep{BMT02}. 
Further, for non-interacting systems one can prove that no shortcut based on invariants or scaling exists in  time-dependent homogeneous potentials.
At the single-particle level, this follows from the fact that the family of trajectories for the width  $\xi(t)$ of a box-like potential  for which a dynamical  invariant exist \citep{boxlaws1}, takes the form 
$\xi(t)=[at^2+bt+c]^{\frac{1}{2}}$, which is incompatible with the boundary conditions required to reduce a time-evolving scaling solution to the initial and target states.
For many-body quantum fluids, the same result is derived from the consistency equations for self-similar dynamics to occur.
To find a shortcut in this scenario one has to relax the condition on the confinement and allow for a inhomogeneous auxiliary harmonic potential 
of the form \citep{DB12}
\beqa
U^{\rm aux}(\x,t)=-\frac{1}{2}m\frac{\ddot{\xi}(t)}{\xi(t)}|\x|^2,
\eeqa
where ${\x}\in\mathbb{R}^D$, $|\x|\in[0,\xi(t)]$. For $D=1$, a box with one stationary wall at $x=0$ and  moving wall at $x=\xi(t)$ is assumed.  Cylindrical and spherical symmetry is imposed for $D=2,3$ respectively. 
This auxiliary potential can be implemented  by means of a blue-detuned laser \citep{Khaykovich} or direct painting with a  rapidly moving laser \citep{prepainters1,prepainters2,painters}. 
Thanks to its presence it is possible to find dynamical self-similar solutions to single-particle and many-body Schr\"odinger equations for time-dependent box-like confinements with a general 
modulation of the width $\xi(t)$. In particular, consider the Hamiltonian
\beqa
\label{mbh}
\mathcal{H}=\!
\sum_{i=1}^N\!\Big[-\frac{\hbar^2}{2m}\Delta_{i}
+U^{\rm aux}(\x_i,t)\!\Big]\!+\!\epsilon\!\sum_{i<j}V(\x_i-\x_j),
\eeqa
where ${\x_i}\in\mathbb{R}^D$, $r_i=|\x|_i\in[0,\xi(t)]$, and let us introduce the scaling factor $\rho(t)=\xi(t)/\xi(0)$.
 If ${\rm V}(\lambda \x)=\lambda^{-\alpha}{\rm V}(\x)$, $\epsilon(t)=\rho(t)^{\alpha -2}$, in the presence of $U^{\rm aux}(\x,t)$, the time evolution of an initial eigenstate of the system with chemical potential $\mu$ follows a scaling law in Eq. (\ref{ad_scaling}). Given the existence of a scaling law, a many-body shortcut can be engineered by designing the scaling factor as for the simple harmonic oscillator, ensuring that the time-evolving state reduces to the initial and target state at the begining and end of the evolution \citep{DB12}. Naturally, this is {possible as well} for Bose-Einstein condensates in the mean-field regime, extending the Castin-Dum-Kagan-Surkov-Shlyapnikov scaling ansatz \citep{DB12}. 

Along a shortcut to an adiabatic expansion, the auxiliary potential is  expulsive in an early stage of the expansion, expelling the atoms from the center and providing the required speed-up.
The rapidly expanding cloud is slowed down in a second state of the expansion, when $U^{\rm aux}(\x,t)$ becomes a trapping potential. 
The sequence is reversed in a shortcut to an adiabatic compression. In both cases, at $t=t_f$,  $U^{\rm aux}(\x,t)$ vanishes, and the cloud reaches the target state, a stationary state of the final Hamiltonian. As a result, STA provide a variant of the paradigmatic model of a quantum piston \citep{QJ12}.
\subsection{Experimental Realization}
Experiments of fast shortcut expansions have been realized 
at Nice with magnetic confining of ${}^{87}$Rb atoms  for ultra-cold clouds
\citep{Nice10} and condensates in the Tomas-Fermi regime \citep{Nice11}.   
Compared to the simple expansions treated in \citep{Ch10}, gravity introduces and extra linear term in the Hamiltonian and requires 
a treatment with additional boundary conditions. 
 
For the cold cloud, samples of $N = 10^5$ atoms and temperature 
$T_0 = 1.63$ $\mu$K were used to keep the time between collisions small $\approx 28$ ms,
and the potential effectively harmonic. The initial trap frequencies for $x,y,z$ directions in Hz were $(228.1, 22.2, 235.8)$ and the final ones $(18.1, 7.1, 15.7)$. The results for the fast (35 ms) 15-fold frequency decompression to the trap in the vertical dimension, yielded a residual center-of-mass oscillation of the cloud equivalent to that of a 1.3-s-long linear decompression, a reduction by a factor of 37.  
 
For the condensate, the number of atoms was $N =1.3\times10^5$ and the initial temperature 
was $T_0 =130$ nK \citep{Nice11}. 
The potential is $U(r, t) =
m\omega_{\pe}^2(t)(x^2 +z^2)+
\frac{1}{2}m\omega_{\pa}^2(t)y^2
+mgz$. 
Initial radial $(x,z)$ and axial $(y)$
frequencies were 235.8 Hz and
22.2 Hz, respectively. The experiment performed a 30-ms-long radial
decompression of the trap by a factor of 9, yielding a
final radial frequency of 26.2 Hz. The axial frequency was
reduced by a factor of 3 to a final value 7.4 Hz. 
Using scaling techniques similar to the ones in Sec. \ref{econd} it was shown that 
this decompression is a shortcut for both directions. 
Residual excitations were attributed to imperfect implementation of $\omega(t)$, 
anharmonicities, and trap tilting.  
\subsection{Optimal Control}
The time-dependent frequency of a
harmonic trap expansion based on invariants can be optimized with respect to
time or to transient excitation energy, restricting the
allowed transient frequencies \citep{Li10,Li11}.

Kosloff and coworkers have applied OCT  to minimize the expansion time with ``frictionless conditions'', i.e., 
taking an initial thermal equilibrium at one temperature into thermal equilibrium at another temperature in a cooling cycle,
using real or imaginary  bang-bang (piecewise constant or ramped) intermediate trap frequencies, see e.g. \citet{Salamon09,EPL11}.  
\subsection{Other Applications}
Inverse engineering expansions using invariant theory or scaling
laws have been applied in several contexts. 
For example,  \citet{Onofrio11}
discussed the possibility of achieving
deep degeneracy of Fermi gases via sympathetic cooling 
by changing the trapping frequency of another species (the coolant)
to keep constant the Lewis-Riesenfeld  invariant. The identified
advantages are the maximal heat capacity
retained by the coolant due to the conservation of the number
of atoms, and the preservation of its phase-space density in
the nondegenerate regime where the specific heat retains
its Dulong-Petit value. The limits of the approach are set by the
transient excitation,  
that should be kept below some allowed threshold, and  
by the spreading of the cooling cloud which reduces the spatial
overlap with the Fermionic cloud. The method is found to be 
quite robust with respect to 
broadband noise in the trapping frequency \citep{Onofrio12}. 
 
\citet{Lianao}  propose a scheme to cool down a mechanical resonator in a three-mirror cavity optomechanical
system. The 
dynamics of the mechanical resonator and cavities is reduced to that of a time-dependent harmonic
oscillator, whose effective frequency can be controlled through the optical driving fields. A simpler harmonic system is studied in \citet{Lianao2}, a charged mechanical resonator coupled to electrodes via Coulomb interaction controlled by bias gate voltages. 

 \citet {PLA2012} designs, using scaling, fast frictionless expansions of an optical lattice with dynamically variable spacing
(accordion lattice). 
Specifically, he considers the 1D Hamiltonian  
$H =p^2/(2m) + V (t) \cos\left(2k_L {x}/{\Lambda(t)}\right)+
{m\omega^2(t)}x^2/2$, where $\Lambda$ is the scale parameter 
which goes from 1 at $t=0$ to $c$ at $t_f$ and 
the parabolic potential only acts during the expansion
according to  
$\omega^2(t)=-\Lambda^{-1}{\partial^2\Lambda}/{\partial t^2}$.
Decreasing the potential depth as $V(t)=V_0/\Lambda^2(t)$, 
and making the first and second derivatives of $\Lambda$ vanish at the boundary times 
guarantee a frictionless expansion. 
In \citet{Yuce2} the results are extended to a continuously replenished BEC
in a harmonic trap or in an optical lattice.  

\citet{simu} propose inverse engineering {of the trap frequencies based on the}  Lewis-Riesenfeld invariants as part of the elementary operations necessary to implement a universal bosonic simulator  using ions in separate traps. This method would allow to improve the accuracy and speed of conventional laser operations {on ions} which are limited by the Lamb-Dicke approximation. 

\citet{JJ} develop a method to  
produce highly coherent-spin-squeezed many-body states in
bosonic Josephson junctions (BJJs). They start from the known mapping of the two-site Bose-Hubbard
(BH) Hamiltonian to that of a single effective particle evolving according to a Schr\"odinger-like
equation in Fock space. Since, for repulsive interactions, the effective potential in Fock space
is nearly parabolic, the inversion protocols for shortcuts to adiabatic evolution in
harmonic potentials may be applied to the many-body BH Hamiltonian.
The procedure
requires a good control of the time variation of the atom-atom
scattering length during the desired period, a possibility
now at hand in current experimental setups for
internal BJJs.
%
%
%
%
%
%
%
%
%
\section{Transport\label{secTRA}}
The efficient transport of atoms and ions by moving the confining trap
is a necessary fundamental requirement for many applications.
These are for example
quantum information processing in multiplexed trap arrays \citep{Leibfried2002,ions,Bowler} or quantum registers \citep{MeschNature}; 
controlled translation from the
production or cooling chamber to {the} interaction or manipulation zones;  
control of interaction times and locations, e.g. in 
cavity QED experiments,
quantum gates \citep{Calarco2000} or metrology \citep{Maleki}; and velocity control to stop \citep{catcher1,boxlaws3} or launch 
atoms \citep{Meschede}. 

The transport should ideally be lossless, fast and ``faithful'', i.e. the final state should be equal to the initial one apart from the translation and possibly phase factors. 
This is compatible with some transient excitation in the instantaneous basis at intermediate times. 

Many different experimental approaches have been implemented. 
Neutral atoms have been transported individually,
as thermal atomic clouds, or  
condensates,  
using optical or magnetic traps.
The magnetic traps 
can be displaced by moving the coils mechanically, by  time-varying currents in a lithographic pattern,
or on a conveyor belt with  permanent magnets \citep{Lahaye}. Optical traps can be used as optical tweezers 
whose focal point is translated by moving mechanically lenses \citep{David}, and traveling lattices (conveyor belts) 
can be made with two counterpropagating beams slightly detuned.
Mixed magneto-optical 
approaches are also possible. To transport ions, controlled time dependent voltages
have been used in linear-trap based frequency standards \citep{Maleki},
and more recently in quantum information applications using multisegmented Paul traps \citep{SK,SK2,Bowler}, or an array of Penning traps \citep{Penning}, also in 2D configurations \citep{Wineland}.

In general, a way to avoid spilling or  excitation of the atoms 
is to perform a sufficiently slow (adiabatic) transport,  
but for many applications the total processing time is limited due to decoherence
and an adiabatic transport may turn out to be too long.
In the context of quantum information processing,
transport could occupy most of the operation time 
of realistic  algorithms, so ``transport times'' need
to be minimized \citep{ions,SK}.
There are in summary important reasons to 
reduce the transport time, and     
several theoretical and experimental works have studied ways to make fast transport 
also faithful \citep{David,Calarco,MNProc,Shan,transport,BECtransport}.
\subsection{Invariant-based Shortcuts for Transport\label{4.1}}
As done for expansions, shortcut techniques can be applied to perform fast atomic transport without final vibrational heating by combining dynamical invariants
and inverse engineering. Two main scenarios can be handled
in this way: shortcuts for the transport of a harmonic trap and  shortcuts
for the transport of an arbitrary trap.
It is also possible to construct shortcuts for more complicated settings like atom stopping or launching, and 
combinations of transport and expansion of harmonic traps.

{\it Transport of a rigid harmonic trap.--}
%
Suppose that a 1D harmonic trap should be moved from $q_0(0)$ at time $t=0$ to $d=q_0(t_f)$ at a time $t_f$. The potential is  
$V=\frac{m}{2} \omega_0^2 (x-q_0(t))^2$ with fixed frequency. Comparing this to  Eq. \eqref{Vinv} this implies 
\beq
F=m\omega_0^2 q_0(t),\; \omega(t)=\omega_0,\;U=0.
\eeq
%
Note
that Eq. \eqref{Erma}
plays no role here and Eq. (\ref{alphaeq}) becomes the only relevant auxiliary equation,
\beq
\label{classical}
\ddot{q}_c+\omega_0^2(q_c-q_0)=0,
\eeq
where $q_c$ can be identified as a classical trajectory. 
This is the equation of a moving oscillator for which an analytical solution is known in both classical and quantum physics.
From a classical mechanics point of view,
the amplitude $\mathcal{A}$ of the oscillatory motion after
transport is the modulus of the Fourier transform of the velocity profile 
associated with 
the trap trajectory  \citep{David}, 
\begin{equation}
\mathcal{A} = |{\mathcal F}[\dot{q}_0](\omega_0)|\,
\label{eq.amplitude}
\end{equation}
with ${\mathcal F}[f]=\int_{-\infty}^{+\infty} f(t){e}^{-{i}\omega
t}\, dt$.
This Fourier formulation of the transport problem allows for many enlightening analogies. For instance,
$\mathcal{A}^2$ is
mathematically identical to the intensity profile for the far
field Fraunhofer diffraction pattern of an object with a
transmittance having the same shape as the velocity profile for
the transport. An optimal transport condition is therefore equivalent to a
dark fringe in the corresponding diffraction pattern. The optimization of the conditions under which a non adiabatic
transport should be carried out with a rigid harmonic trap are thus equivalent to apodization problems in optics. If the velocity profile contains
the repetition of a pattern one expects an interference-like effect, this would be, for instance, the case for a
symmetrical round trip transport as experimentally demonstrated in \citet{David}.

Quantum mechanically, the wave function after transport reads
\begin{equation}
\Psi(q,t_f) = \tilde{\Phi}(q-q_0(t),t)\exp\left(
\frac{im(q-q_0(t))\dot{q}_0}{\hbar}\right)\exp\left(
\frac{i}{\hbar}\int_0^tdt'{\mathcal L}(t')\right),
\label{psiexact}
\end{equation}
where ${\mathcal L}=m{\dot q}_c^2/2-m\omega_0^2(q_c-q_0(t))^2/2$ is the Lagrangian associated with the equation of motion (\ref{classical}), and 
$\tilde{\Phi}$ a wave function that coincides with the initial wave function at initial time and that evolves under the action of the static harmonic potential of angular frequency $\omega_0$ located at $q=q_0(0)$. Using the boundary conditions associated with the transport, one finds from Eq.~(\ref{psiexact}) that an optimal transport for which the system starts in the ground state and ends up in the ground state of the displaced potential corresponds exactly to the classical criterion of a cancellation of the Fourier transform of the velocity profile, i.e. ${\mathcal A} =0$.

%
%
%

Let us now address the application of 
invariant-based engineering. 
We first design an appropriate classical trajectory $q_c(t)$ fulfilling the boundary conditions 
$q_c(0)=q_0(0)=0,\; \dot{q}_c(0)=0,\; \ddot{q}_c(0)=0$ and $q_c(t_f)=q_0(t_f)=d,\; \dot{q}_c(t_f)=0,\; \ddot{q}_c(t_f)=0$, 
to ensure an evolution from the n-th state of the initial trap to the n-th state 
of the final trap. Then the trap motion trajectory $q_0(t)$ is deduced via Eq. (\ref{classical}).
Some variants are vertical transport with a gravity force, so that
$F=m\omega_0^2q_0-mg$ and  Eq. (\ref{classical}) becomes
%
$
\ddot{q}_c+\omega_0^2(q_c-q_0)=-g, 
$
%
%
%
%
and stopping or launching processes \citep{transport}.

A major concern in practice for all these applications is to keep the
harmonic approximation valid. This may require an analysis of the actual potential and of the excitations taking place along the non-adiabatic transport process. Without such detailed analysis, the feasibility of the
approach for a given transport objective set by the pair $d,t_f$ can be estimated by comparing lower excitation bounds \citep{BECtransport}. These are obtained using calculus of variations as we have discussed before for expansions.
Writing the expectation value of potential energy for a transport mode as  $\la V(t)\ra=\frac{\hbar\omega_0}{2}\left(n+1/2\right)+E_P$, 
the time average of $E_P$ is bounded as 
%
${\overline{E_{P}}}\ge{6md^2}/({t_f^4 \omega_0^2})$ \citep{transport}.

{This scaling should be compared to }the milder dependence on $t_f^{-2}$ of the time-averaged transient energy in expansions \citep{energy}. \citet{OCTtrans} have shown how to realise this bound by allowing the discontinuous acceleration of the trap at $t = 0$ and $t = t_f$
and also finite jumps
in the trap position.

In \citet{OCTtrans}, the invariant-based method is complemented
by optimal control theory. Since actual traps are not really
harmonic, the relative
displacement between the center of mass and the trap center
is kept bounded as a  constraint. The trajectories 
are then optimized according to
different physical criteria: time minimization, (time-averaged)
displacement minimization, and (time-averaged) transient
energy minimization. The minimum time solution has a ``bang-bang'' form, and the minimum displacement solution
is of ``bang-off-bang'' form.
In this framework 
discontinuities in the acceleration $\ddot{q}_c$ at the edge times and 
elsewhere are allowed. Physically this means that the trap 
may ideally
jump suddenly over a finite distance, whereas the
velocity $\dot{q}_c$ and the trajectory $q_c$ remain always continuous.

{\it Transport of an arbitrary trap with compensating force.--}
In the second main scenario, the trap potential $U(q-q_0(t))$ is arbitrary, and it is rigidly displaced along $q_0(t)$.
Now, in Eq. (\ref{Vinv}),
%
%
%
$\omega=\omega_0=0$, $F=m\ddot{q}_0$,
and $q_c$ in Eq.~(\ref{alphaeq}) may be identified with the transport function $q_0$. 
Inverse engineering in this case is based on designing the trap trajectory $q_0$ \citep{transport}.
In addition to $U$,
there is a  compensating linear potential term $-mq\ddot{q}_0$ in 
$
H=p^2/2m-mq\ddot{q}_0+U(q-q_0).
$
The corresponding force compensates for the inertial force due to the trap motion in the rest frame of the trap, in such a way that the wave function in that frame is not modified up to a time dependent global phase factor.
This Hamiltonian was originally proposed by \citet{MNProc} using the  ``fast-forward'' scaling technique. \citet{Masuda2012} has recently generalized 
this result for interacting, identical, spinless particles. 
\subsection{Transport of a Bose-Einstein Condensate}
The two main scenarios of the previous subsection can  be generalized
for Bose-Einstein condensates \citep{BECtransport}.
We first consider 1D  harmonic transport.
For the GPE 
\beqa
i\hbar\frac{\partial\psi}{\partial
  t} (q,t)=\left[-\frac{\hbar^2}{2m}\frac{\partial^2}{\partial q^2}+\frac{m\omega_{0}^{2}}{2}(q-q_{0}(t))^2
+ g_{\rm 1}|\psi (q,t)|^2\right]\psi (q,t),
\label{GP2} 
\eeqa
the results of Sec. \ref{4.1} motivate the ansatz
\beq
\psi(q,t)=\exp\left\{\frac{i}{\hbar}\left(-\mu
t+m\dot{q}_c q\right)
- \frac{i}{\hbar} \int_{0}^{t}\!dt'\left[\frac{m}{2}\bigg(\dot{q}_c^2-\omega_0^2(q_c^2-q_0^2)\bigg)\right]\right\}
\chi(\sigma),
\label{trans} 
\eeq
where 
$\chi(\sigma)$ satisfies the stationary GPE
\beqa
\left[-\frac{\hbar^2}{2m}\nabla^2_{{\sigma}}+
\frac{m\omega^{2}_{0}}{2}|{\sigma}|^2+U(\sigma)
+g_{\rm 1}|\chi(\sigma)|^2\right]\chi(\sigma)= \mu\;\chi(\sigma). 
\label{GPstationary}
\eeqa
The ansatz provides indeed 
a solution to Eq. (\ref{GP2}) when   
$q_c(t)$ satisfies Eq. (\ref{classical}).  
Inverse engineering gives the trap  trajectory $q_0 (t)$ from  \eqref{classical} after designing $q_c(t)$, as for the linear dynamics.

The inverse method can also be applied to anharmonic transport of condensates by means of a
compensating force \citep{transport}. 
In either scenario this method does not require that $t_f$ satisfies any 
discretization condition, as it occurs with other approaches 
\citep{BECtransport}, and $t_f$ can in principle be made as small as desired.    
In practice 
there are of course technical and fundamental limitations 
\citep{transport}.
Smaller values of $t_f$ increase the distance from the condensate to the trap center, 
and the effect of anharmonicity. There could be also geometrical constraints: for short $t_f$, $q_0(t)$ 
could exceed the interval [$0,d$]. OCT combined 
with the inverse method, see below, provides a way to design 
trajectories taking these restrictions into account.  
{\it Optimal control theory.---}
An OCT trajectory has been found 
when the center of the
physical trap is kept inside a given range (e.g. inside the vacuum
chamber), i.e.  $q_{\downarrow} \le q_0 (t) \le q_{\uparrow}$ \citep{BECtransport}. 
At the beginning the trap is immediately
set at the upper bound $q_{\uparrow}$ to accelerate the condensate as
much as possible and at time $t_1$ the trap is moved to the lower bound
$q_{\downarrow} $ to decelerate the condensate so as to leave it  at rest at  $t_f$. 
An important open question 
is to evaluate 
the effect of the approximate realization of the 
discontinuities found in the bang-bang solutions.

{\it Effect of Perturbations.--}
\citet{BECtransport} also investigated the effect of 
anharmonicities when the harmonic transport protocol is applied. For a symmetrically perturbed
potential
%
$
V=\omega_0^2m \left[(q-q_0)^2 + \alpha (q-q_0)^4\right]/2,
$
the fidelity increases with increasing coupling constant $g_{\rm 1}$, because of the increased width of the wavefunction.   
They also considered that the center of the physical trap is 
randomly perturbed with respect to 
$q_0(t)$.
The fidelity at $t_f$ is found to be 
independent of $d$ and the chosen $q_c(t)$
and increases for shorter times $t_f$ and   
for smaller couplings $g_1$,
unlike the previous results.
\section{Internal State engineering\label{secINT}}
Manipulating the internal state of a quantum system with time-dependent interacting
fields is the basis of quantum information
processing \citep{Allen,Vitanov-Rev,Bergmann} and many other fields.
Two major routes 
are resonant pulses, 
and adiabatic methods such as ``Rapid'' Adiabatic Passage (RAP), Stimulated Raman Adiabatic Passage (STIRAP), and their variants. Simple fixed-area resonant pulses, such as a $\pi$ pulse, may be fast if intense enough, but 
they are also highly sensitive to 
variations in the pulse area, and to inhomogeneities in the sample \citep{Allen}.
Composite pulses provide an alternative to the single $\pi$ pulse, 
with some successful applications \citep{Levitt,Collin,Torosov}, 
but still {they} need an accurate control of pulse phase and intensity.
In NMR, composite pulses are being superseded by adiabatic passage methods, which have also been very successful in laser cooling, chemical reaction dynamics, metrology, 
atom optics, interferometry, or cavity quantum electrodynamics.
Adiabatic passage is robust versus 
parameter variations but slow. 
It is moreover  
prone to decoherence because of the effect of noise  
over the long times required. 
This motivates the search for fast and robust shortcuts, with respect to parameter variations and noise.

Several methods to find  
STA  have
been put forward 
for two- and three-level atomic systems.
Among them,  methods that we have  
already discussed in Sec. 2, like the transitionless driving, invariant-based engineering, 
or OCT. 
\subsection{Population Inversion in Two-level Systems\label{poin}}
%
%
%
Using the convention $|1\ra={1\choose 0}$, $|2\ra={0\choose 1}$,
%
%
assume a two-level system with a Hamiltonian of the form
\beqa
H_0 (t)= \frac{\hbar}{2} \left(\begin{array}{cc} -\Delta(t) & \Omega_{R}(t) -i\Omega_I(t)
\\
\Omega_{R}(t)+i\Omega_I(t) &  \Delta(t)
\end{array}\right).
\label{H0}
\eeqa
In quantum optics it describes the semiclassical 
coupling of two atomic levels with a laser in a laser-adapted interaction picture,
where $\Omega_c(t)=\Omega_R(t) + i \Omega_I(t)$ is the complex Rabi
frequency and $\Delta(t)$ the time-dependent detuning between laser and transition
frequencies. We will keep the language of the atom-laser interaction
hereafter, but  in other two-level systems, 
for example, in a spin-$1/2$ system or in a Bose-Einstein condensate on an 
accelerated optical lattice \citep{Oliver},
$\Omega_c(t)$ and $\Delta(t)$ may 
correspond to different physical quantities.
     
Initially at time $t=0$, the atom is assumed to be in the ground state $|1\ra$.
The goal is to
achieve a perfect population inversion such that at a time $t=T$ the atom
is in the excited state.
For a $\pi$ pulse 
the laser is on resonance, i.e. $\Delta(t)=0$ for
all $t$. If the Rabi frequency is chosen like $\Omega_c(t) = \fabs{\Omega_c (t)}
e^{i\alpha}$, with a time-independent $\alpha$, and such that
%
$
\int_0^T dt\, \fabs{\Omega_c(t)} = \pi, 
$
%
the population is inverted at time $T$.
A simple example is the ``flat'' $\pi$ pulse with $\Omega_c(t) = e^{i\alpha}{\pi }/{T}$.

Adiabatic schemes provide another major route for population inversion.
In the ``Rapid Adiabatic Passage'' technique 
the radiation is swept slowly through resonance.
The term ``rapid'' here means that the frequency sweep is shorter than the life-time of 
spontaneous emission and other relaxation times.
Many schemes corresponding to different functions $\Delta(t)$, and $\Omega_R(t)$ are possible.
The simplest is a Landau-Zener approach, with $\Delta$ linear in time, $\Omega_R$ constant and $\Omega_I=0$.

{\it Transitionless shortcuts to adiabaticity.--}
If an adiabatic scheme is used and the adiabaticity condition $\frac{1}{2}|\Omega_a|\ll
|\Omega(t)|$ (where $\Omega=
\sqrt{\Delta^2+\Omega_R^2}$, $\Omega_I=0$, and  $\Omega_a \equiv [\Omega_R \dot{\Delta} - \dot{\Omega}_R \Delta]/\Omega^2$)
is not fulfilled, the inversion fails. 
We may still get an inversion (i.e. a shortcut)
by applying a counterdiabatic field 
such that its maximum is not larger than the maximum of $\Omega_R$ \citep{Ch10b}.    
%
The total Hamiltonian, see Sec. \ref{secCD}, for the transitionless shortcut protocol is 
\citep{Rice08,Berry09,Ch10b}
\beqa
H_{0a} (t)= \frac{\hbar}{2} \left(\begin{array}{cc} -\Delta & \Omega_{R}- i  \Omega_a
\\
\Omega_{R}+ i  \Omega_a &  \Delta
\end{array}\right). 
\label{toha}
\eeqa

{\it Invariant-based Shortcuts.--}
STA in two-level systems can be also found making 
use of Lewis-Riesenfeld
invariants \citep{Yidun,noise}. 
For $H_0$ in Eq. (\ref{H0}), 
a dynamical invariant may be parameterized as 
\beqa
\label{I}
I (t)= \frac{\hbar}{2} \mu \left(\begin{array}{cc} \fcos{\Theta(t)} & \fsin{\Theta(t)} e^{- i \alpha(t)}
\\ \fsin{\Theta(t)} e^{i \alpha(t)} &  -\fcos{\Theta(t)}
\end{array}\right),
\eeqa
where $\mu$ is a constant with units of frequency to keep $I(t)$ with dimensions of energy. From the invariance condition the functions $\Theta(t)$ and $\alpha(t)$
must satisfy 
\begin{eqnarray}
\begin{array}{l}
\dot\Theta = \Omega_I \cos \alpha - \Omega_R \sin \alpha,\\
\dot\alpha =  -\Delta(t) - \cot\Theta\left(\Omega_R\cos\alpha  +
\Omega_I\sin\alpha\right).
\end{array}
\label{schrpure}
\end{eqnarray}
The eigenvectors of the invariant are
\beqa
|\phi_+(t)\ra = \left(
\begin{array}{c} 
\fcos{\Theta/2} e^{-i\alpha/2}\\
\fsin{\Theta/2}e^{i\alpha/2}
\end{array}
\right), \quad
|\phi_-(t)\ra = \left(
\begin{array}{c}
\fsin{\Theta/2} e^{-i\alpha/2}\\
-\fcos{\Theta/2} e^{i\alpha/2}
\end{array}
\right),
\label{phiplus}
\eeqa
with  eigenvalues $\pm \frac{\hbar}{2} \mu$.
A general solution $|\Psi(t)\ra$ of the Schr\"odinger equation can be written as a linear
combination 
\begin{eqnarray}
|\Psi(t)\ra = c_+ e^{i\kappa_+(t)} |\phi_+(t)\ra + c_- e^{i\kappa_-(t)} |\phi_-(t)\ra,
\end{eqnarray}
where $c_\pm$ are complex, constant coefficients, and 
$\kappa_\pm$ are the phases of \citet{LR} {introduced in Eq. \eqref{LRphase}}.
Let 
$
\gamma = - 2 \kappa_+ = 2 \kappa_-,
$
then $\gamma$ must be a solution of
\begin{eqnarray}
\dot\gamma &=& \frac{1}{\sin\Theta} \left(\cos\alpha\,\Omega_R +
\sin\alpha\,\Omega_I\right).
\label{dotgamma}
\end{eqnarray}
Equivalently a solution of the Schr\"odinger equation
$|\Psi(t)\ra$ may be designed  
with the same parameterization as above
($|\Psi(t)\ra\la\Psi(t)|$ is a dynamical invariant.)
and, by putting this ansatz into the Schr\"odinger equation, 
Eqs. (\ref{schrpure}) and (\ref{dotgamma}) are found. 

If $\Omega_R(t)$, $\Omega_I(t)$ and $\Delta(t)$ are given,
Eqs. \eqref{schrpure} and \eqref{dotgamma} have to be solved to get
$\Theta(t)$, $\alpha(t)$ and $\gamma(t)$. A particular solution of the
Schr\"odinger equation is then given by
\begin{eqnarray}
|\psi(t)\ra = |\phi_+(t)\ra e^{-i \gamma(t)/2}.
\label{solpsi}
\end{eqnarray}
To find invariant-based shortcuts and  inverse
engineer the Hamiltonian 
$\Theta(t)$, $\alpha(t)$, and $\gamma(t)$ are fixed first, fulfilling the 
boundary conditions $\Theta(0)=0$ and $\Theta(T)=\pi$. The 
wave function \eqref{solpsi} corresponds then to an atom in the ground state at $t=0$ and in the excited state at $t=T$, i.e. a perfect population inversion.
Then, by inverting
\eqref{schrpure} and \eqref{dotgamma}, 
\begin{eqnarray}
\Omega_R &=& \cos\alpha\sin\Theta \;\dot\gamma -
\sin\alpha\;\dot\Theta\label{pot_R},
\\
\Omega_I &=& \sin\alpha\sin\Theta\;\dot\gamma +
\cos\alpha\;\dot\Theta\label{pot_I},
\\
\Delta &=& -\cos\Theta \;\dot\gamma - \dot\alpha.
\label{pot_D}
\end{eqnarray}
There is much freedom in designing such a shortcut because the auxiliary functions $\Theta(t)$, $\alpha(t)$ and $\gamma(t)$ can be chosen arbitrarily
except for the boundary conditions.
\subsection{Effect of Noise and Perturbations}
A key aspect to choose among the many possible shortcuts is their stability or robustness versus different perturbations. 
\citet{noise} have derived optimal invariant-based shortcut protocols, maximally stable concerning amplitude noise of the interaction and with respect to systematic errors. 
It turns out that the perturbations due to noise and systematic errors require different optimal protocols. 

Let the ideal, unperturbed Hamiltonian be the $H_0(t)$ of Eq. \eqref{H0}. 
In \citet{noise}, it is assumed that the errors affect $\Omega_{R}$
and $\Omega_{I}$ 
but not the detuning $\Delta$, which, for an atom-laser 
realization of the two-level system is more easily controlled. 

For systematic errors, for example if different atoms at different positions are subjected
to slightly different fields due to the Gaussian shape of the
laser, 
the actual, experimentally implemented 
Hamiltonian is $H_{01}=H_0
+ \beta H_1$, where
$
H_1 (t)= 
H_0 (t)\big|_{\Delta\equiv 0} 
$
and $\beta$ is the amplitude of the systematic error.

The second type of error considered in \citet{noise} 
is amplitude noise, which is assumed to affect $\Omega_R$
and $\Omega_I$ independently with the same strength parameter $\lambda^2$. This is motivated
by the assumption that two lasers may be used to implement the two parts
of the Rabi frequency.
The final master equation describing systematic error and amplitude-noise error
is
\begin{eqnarray}
 \frac{d}{dt} \hat\rho &=& -\frac{i}{\hbar} [H_0 + \beta H_1,\hat\rho]
 -\frac{\lambda^2}{2 \hbar^2} \left([H_{2R},[H_{2R},\hat\rho]] +
     [H_{2I},[H_{2I},\hat\rho]]\right),
\label{masterfinal}
\end{eqnarray}
where
%
$H_{2R} (t)= H_0 (t)\big|_{\Delta\equiv\Omega_I\equiv 0}$,
and  
$H_{2I} (t)= H_0 (t)\big|_{\Delta\equiv\Omega_R\equiv 0}$.

Before studying both types of error together it is fruitful to look
at them separately.

{\it Amplitude-Noise Error.--}
If there is no systematic error ($\beta=0$) and only
an amplitude-noise error 
affecting the Rabi frequencies,  a noise sensitivity can be defined as
\begin{eqnarray*}
q_N := -\frac{1}{2} \left. \frac{\partial^2 P_2}{\partial \lambda^2}\right|_{\lambda=0}
= - \left.\frac{\partial P_2}{\partial (\lambda^2)}\right|_{\lambda=0},
\end{eqnarray*}
where $P_2$ is the probability to be in the excited state at final time $T$,
i.e. $P_2 \approx 1 - q_N \lambda^2$.

To find an invariant-based shortcut protocol maximally stable concerning amplitude noise,
it is first assumed that
the unperturbed solution (\ref{solpsi})
satisfies $\Theta(0)=0$
and $\Theta(T)=\pi$. Using a perturbation approximation  of the
solution and keeping only terms up to $\lambda^2$ \citep{noise}, 
\begin{eqnarray}
q_N &=& \frac{1}{4} \int_0^T dt \Big[
(\cos^2\Theta + \cos^2\alpha\sin^2\Theta)(m\sin\alpha - \cos\alpha
    \dot\Theta)^2\nonumber\\
 && + (\cos^2\Theta + \sin^2\alpha\sin^2\Theta)(m\cos\alpha + \sin\alpha
    \dot\Theta)^2\Big],
\end{eqnarray}
where $m(t)=-\dot\gamma \sin\Theta$.
Minimizing the error sensitivity $q_N$ by Euler-Lagrange
one gets that the optimal solutions satisfy \citep{noise}  $\alpha=n \pi/4$, $n$ odd, and
\begin{eqnarray}
(3 + \cos(2\Theta))\ddot\Theta = \sin(2\Theta) (\dot\Theta)^2.
\label{eqx}
\end{eqnarray}
The corresponding $\Omega_R$ and $\Omega_I$
can be calculated from Eqs. (\ref{pot_R}) and (\ref{pot_I}).
In this case, $\Omega_{R} = \pm \dot\Theta/\sqrt{2} = \pm \Omega_I$ and $\Delta(t)=0$. 
The optimal noise sensitivity value is $q_N = 1.82424/T< \pi^2/(4T)$
and the maximum of the Rabi frequency is $\Omega_R(t_f/2) t_f \approx
 2.70129$.
An approximate solution of Eq. (\ref{eqx}) is given by $\Theta(t) = \pi t/T -
\frac{1}{12}\sin(2\pi t/T)$, with a noise sensitivity of $q_N=1.82538/T$.

{\it Systematic Error.---}
If there is no amplitude-noise error ($\lambda=0$) and only
systematic error, a systematic error
sensitivity is defined as 
\begin{eqnarray*}
q_S := -\frac{1}{2}\left. \frac{\partial^2 P_2}{\partial \beta^2}
\right|_{\beta=0} = -\left. \frac{\partial P_2}{\partial (\beta^2)}
\right|_{\beta=0},
\end{eqnarray*}
where $P_2$ is as before the probability to find the atom in the excited
state at final time
$T$. $q_S$ may be calculated with a 
perturbation approximation of the
solution keeping only terms up to $\beta^2$
\citep{noise}.

To find an optimal scheme the invariant based technique
is used again. The evolution of the unperturbed state
can be parameterized as 
before, $|\psi(t)\ra$ (see Eq.~\eqref{solpsi}), 
with the boundary values  $\Theta(0)=0$ and $\Theta(T)=\pi$.
The expression for the systematic error sensitivity is now
\begin{eqnarray*}
q_S = \left|\int_0^T dt e^{-i\gamma}\dot\Theta \sin^2\Theta\right|^2.
\end{eqnarray*}
The optimal value is clearly $q_S = 0$. An example of
a class which fulfills $q_S=0$ is found by 
letting
$
\gamma(t) = n \left(2 \Theta - \sin(2\Theta)\right).
$
If follows that 
$
q_S = {\sin^2\left(n\pi\right)}/(4n^2), 
$
so for $n=1,2,3,...$, 
$q_S = 0$.
There is still some freedom left, this allows further optimization
concerning additional constraints. 

%

\begin{figure}[htbp]
\begin{center}
\includegraphics[width=1\linewidth]{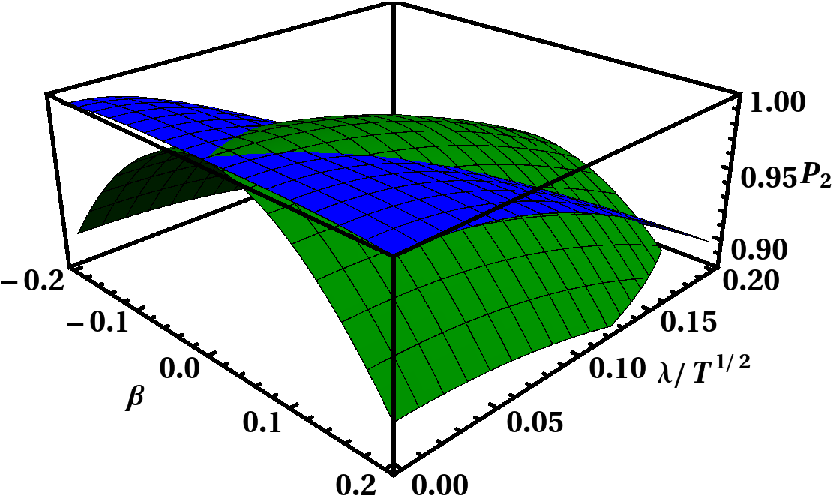}
\caption{{\bf (Color online) Probability $P_2$ versus noise error and systematic
error parameter;  optimal systematic stability protocol (blue), optimal noise protocol (green).}}
\label{general_ex}
\end{center}
\end{figure}

{\it Systematic and amplitude-noise errors.--}
If both errors coexist the
optimal schemes will depend on their relative importance. \citet{noise}
examine numerically the
behavior of different protocols. 
Fig. \ref{general_ex} shows that the different optimal schemes perform better than the other one 
depending on the dominance of one or the other type of error. 
%

%
%

Additional work is in required  to 
extend the results in \citet{noise}
to different types of noise and perturbations. 
Apart from the invariant-based approach, \citet{Lacour} have  
proposed robust trajectories in the adiabatic parameter space that maximize the population transfer  for a two-level system subjected to dephasing. Also the robustness of the ``parallel adiabatic passage'' technique (keeping the eigenvalues of the Hamiltonian parallel \citep{PLAP3}) with respect to fluctuations of the phase, amplitude and pulse area was analyzed in \citet{PLAP4}.  
\subsection{Three-level Systems}
The transitionless driving for stimulated rapid adiabatic passage from level 1 to level 3 in a lambda configuration with an intermediate state 2
making use of a pumping and a Stokes laser   
was studied in \citet{ShoreOC,Rice03,Rice05,Ch10b}.   
The fast-driving cd field connects levels $|1\ra$ and $|3\ra$.
This implies in general a weak magnetic dipole transition, which limits
the ability of the field to shorten the times. 
Invariant-based engineering solves the problem  by providing
alternative shortcuts that do not 
couple directly levels $|1\ra$ and $|3\ra$ \citep{Chen3}, as discussed below. 
It should be noted though that 
in an optical analogy of STA to engineer multimode waveguides all these schemes 
(with or without 1-3 coupling) may in principle be implemented \citep{SHAPEapp,Tseng}  by computer-generated holograms. In this analogy, based on the paraxial approximation, space plays the role of time so that the effect of the shortcuts is to shorten the length of the mode converters.      

In \citet{Chen3}, using two lasers on resonance
with the 1-2 and 2-3 transitions,
two single-mode protocols that make use of one eigenstate of the invariant
are described.  
In these protocols full fidelity requires an infinite laser intensity, and  
shortening the time also implies an energy cost.
The first protocol, based on simple sine and cosine functions for the pumping and Stokes lasers keeps the population of level 2 small.
To achieve the same fidelity, less intensity is required in the second protocol, 
in which the intermediate level $|2\ra$ is populated. The population of the intermediate level is usually problematic when its time decay
scale is smaller than the process time. While this may be a serious drawback for an adiabatic
slow process, it need not be for a fast shortcut. Protocols that populate level $2$ may thus be considered as useful alternatives for certain systems and sufficiently short process times.
   
In the previous two protocols the initial state is not exactly $|1\ra$ to avoid a divergence in the Rabi frequency. 
A third multi-mode wave-function protocol is also proposed in \citet{Chen3} using the same fields as for the first protocol but with an initial state which is simply the bare state $|1\ra$. 
It provides a much less costly shortcut so exploring the multi-mode approach for this and other systems is an interesting task for future work. 

Inverse engineering of four-level systems has been considered
in \citet{Yidun1}, where a full Lie-algebraic classiffication and
detailed construction of the dynamical invariants is provided.  
\subsection{Spintronics}
Coherent spin manipulation in quantum dots is the key element in the state-of-the-art technology of spintronics. 
\citet{Ban} have considered the electric control of electron spin in a quantum dot formed in a two-dimensional electron gas confined by the material composition under a weak magnetic field, focusing on the spin flip in the doublet of the lowest orbital state. The influence from higher orbital states can be taken into account by the L\"{o}wdin partition technique
reducing the full Hamiltonian into an effective two-level one in which the 
matrix elements depend on electric field components. 
Using invariant-based inverse engineering the time-dependent electric fields are designed so to flip the spin rapidly and avoid decoherence effects. The results are stable with respect to environmental noise and the device-dependent noise and may open  
new possibilities for high-fidelity spin-based quantum information processing.
\subsection{Experiments}
The counterdiabatic or transitionless approach described in Secs. \ref{secCD}, 2.4, and 
\ref{poin} has been applied recently to invert the population of 
different two-level 
systems: 

In \cite{Oliver} the effective two-level system is set as a condensate in the bands of an accelerated optical lattice \cite{Oliver0}. Writting the Hamiltonian in Cartesian-like 
coordinates as $H=X\sigma_x+Y\sigma_y+Z\sigma_z$,
$X$ may be controlled by the trap depth, $Z$ by the
lattice acceleration \cite{Oliver0}, and $Y$ could in principle
be implemented by a second shifted lattice. The counterdiabatic term 
in Eqs. (\ref{28}) or (\ref{toha}) is of the form  $Y\sigma_y$ 
whose realization is cumbersome in this setting.  The alternative was to perform a unitary transformation that leads to the same final state  modifying the original
$X$ and $Z$ terms. This manipulation, discussed in Sec. 2.4. (see Eq. (\ref{u3}), 
was interpreted as a $Z$-rotation in  
\cite{Sara12}, where it is compared to the one based on 
the first-order superadiabatic cd-term $H_{cd}^{(1)}$.\footnote{The use of the term ``superadiabatic'' in \cite{Oliver} differs -is broader there- from  the
one in Sec. 2.4.} Landau-Zener and a ``tangent'' protocol with a tangent function for $Z$
unnafected by the rotation, are used as a reference, 
the later being found to be very robust versus a simulated
variation of control parameters.  

In \cite{expcd} the two-level system is  a single nitrogen vacancy center in diamond
controlled by time dependent microwave fields.
The reference process is a Landau-Zener transition, and the $Y\sigma_y$ cd-term 
is implemented by a field oscillating $\pi/2$ radians out of phase with respect to 
the field that provides the $X\sigma_x$ term. As the maximal value of the total field amplitude is 
bounded, 
in this case to avoid undesired transitions, a ``rapid-scan'' approach is
implemented to shorten the protocol time: the protocol is divided into a discrete set of time 
segments with varying phase and the time duration of each segment is adjusted so that 
the maximal amplitude allowed is applied.  
\section{Wavepacket Splitting}
%
%
%
%
%
Splitting a wavefunction  
without excitating it is important in matter wave interferometry \citep{S07,S09a,S09b,Augusto}.
For linear waves, described by the Schr\"odinger equation, it is a peculiar operation, as adiabatic following is not  robust but  unstable with respect to a small external potential asymmetry \citep{JGB,split}. The ground-state wavefunction ``collapses'' into the slightly lower well so that a very slow trap potential bifurcation fails to split the wave except 
for perfectly symmetrical potentials.  
A fast bifurcation  
with a rapidly growing separating potential 
succeeds to split the wave but at the price of
a strong excitation. 
STA that speed up the adiabatic process along a non-adiabatic route overcome these problems \citep{split}.  
Numerical modelling shows that the wave splitting via shortcuts is significantly more stable than the adiabatic following with respect to  asymmetric perturbations and avoids the final excitation. 
Specifically  \citet{split} use the  streamlined version \citep{ErikFF} of the fast-forward technique of \citet{MNProc}, see Sec. \ref{secmethod}, applied to Gross-Pitaevskii  or Schr\"odinger equations  
after having found some obstacles to apply 
the invariants-based method (the eigenvectors of quadratic-in momentum invariants do not satisfy the required boundary conditions \citep{ErikFF}), and the transitionless-driving algorithm \citep{Rice03} (because of difficulties to implement in practice the counter-diabatic terms).
The following discussion refers to the Schr\"odinger equation except for a final comment on the GPE.  

{\it{Fast-forward approach.}---}
To apply the FF approach the density $r(x,t)$ must be first designed. 
Assume the splitting of an initial single Gaussian 
$f(x,0)=e^{-x^2/(2 a_0^2)}$, where $a_0$ is the width
of the ground state for a harmonic oscillator with frequency $\omega/\hbar$, $a_0=\sqrt{\hbar/(m\omega)}$,  
into a final double Gaussian $f(x,t_f)=e^{-(x-x_f)^2/(2a_0^2)}+e^{-(x+x_f)^2/(2a_0^2)}$. 
The interpolation 
\beq
\label{ansatzrbueno}
r(x,t)=z(t)[e^{- (x-x_f(t))^2/(2a_0^2)}+e^{- (x+x_f(t))^2/(2a_0^2)}],
\eeq
where $z(t)$ is a normalization function,
generates simple $Y$-shaped potentials. 
The conditions $\dot{x}_0(0)=\dot{x}_0(t_f)=0$ are imposed, so $\dot r=0$ at  boundary times.
In \citet{split} $x_0(s)=x_f(3s^2-2 s^3)$, where $s=t/t_f$, is chosen 
for simplicity,   
and Eq. (\ref{imag0}) is solved for the initial conditions
to get the FF potential with Eq. (\ref{real}). 
%
%

{\it{Effect of the perturbation}.---}
The effects of an asymmetric perturbation may be studied 
with the 
potential $V_{\lambda}=V_{FF}+\lambda\theta(x)$,
where $\theta$ is the step function and $V_{FF}$ the potential obtained via 
Eqs. (\ref{real}), (\ref{imag0}), and (\ref{ansatzrbueno}) with $\lambda=0$.  
The goal is to find a stable time-dependent porotocol that, even without knowing the value of $\lambda$, is able to produce the split state.

{\it{Moving two-mode model}.---}  
Static two-mode models have been  
used before     
to analyze splitting processes \citep{Javanainen99,S09b,Aichmayr}. 
\citet{split} consider instead a  
two-level model with moving left and right basis functions to 
provide
analytical estimates and insight as a complement of the 
more detailed FF approach. 

 
Assume first the 
(symmetrical and orthogonal) moving left and right bare basis states 
$|L(t)\rangle = {0 \choose 1}$, $|R(t)\rangle ={1\choose 0}$, 
and a corresponding two-mode Hamiltonian model 
\beq
\label{H_tm}
H(t)=\frac{1}{2} \left ( \begin{array}{cc}
\lambda
& -\delta(t)\\
-\delta(t)& -\lambda
\end{array} \right),
\eeq 
where $\delta(t)/\hbar$ is the tunneling rate \citep{Javanainen99,S09b} and $\lambda$ the  energy difference between the two wells \citep{Aichmayr}.
We may simply consider $\lambda$ constant through a given splitting process  and equal to the perturbative parameter that defines the asymmetry.  
Thus, the instantaneous eigenvalues are $E^{\pm}_\lambda(t)=\pm \frac{1}{2} \sqrt{\lambda^2+\delta^2(t)}$, and the normalized eigenstates 
\be
\begin{array}{ll}
\label{eigenstates_tm}
|\psi^+_\lambda(t)\rangle = \sin{ \left ( \frac{\alpha}{2} \right ) } |L(t)\rangle-\cos {\left ( \frac{\alpha}{2} \right ) }|R(t)\rangle, 
\\
|\psi^-_\lambda(t)\rangle = \cos{ \left ( \frac{\alpha}{2} \right )}|L(t)\rangle+\sin{\left ( \frac{\alpha}{2} \right )}|R(t)\rangle,
\end{array}
\eeq
where $\tan \alpha = \delta (t)/\lambda$ defines the mixing angle.

When $\left \{ |L(t)\rangle,|R(t)\rangle \right\}$ 
are close enough initially (and $\delta(0)\gg\lambda$), the instantaneous eigenstates of $H$ are close to the symmetric ground state 
$|\psi^{-}_{0}(0)\rangle=\frac{1}{\sqrt{2}}(|L(0)\rangle+|R(0)\rangle)$
and the antisymmetric excited state 
$|\psi^{+}_{0}(0)\rangle=\frac{1}{\sqrt{2}}(|L(0)\rangle-|R(0)\rangle)$
of the single well.
At $t_f$ two extreme regimes may be distinguished:
{\it{i)}} For $\delta(t_f)\gg\lambda$ the final eigenstates of $H$ tend to $|\psi^{\mp}_{\lambda}(t_f)\rangle=\frac{1}{\sqrt{2}}(|L(t_f)\rangle\pm
|R(t_f)\rangle)$ which correspond to the symmetric and antisymmetric splitting states. 
{\it{ii)}} For  $\delta(t_f)\ll\lambda$ the final eigenfunctions of
$H$ collapse and become right and left localized states: $|\psi^{-}_{\lambda}(t_f)\rangle=|L(t_f)\rangle$ and $|\psi^{+}_{\lambda}(t_f)\rangle=|R(t_f)\rangle$.
Since $\delta(t_f)$ is set as a small number to avoid tunnelling
in the final configuration, 
the transition from one to the other regime explains the collapse 
of the ground state function to one of the wells at small 
$\lambda\approx\delta(t_f)$.     

{\it{Dynamics of the two-mode model}.---}
In a moving-frame interaction-picture wave function  
$\psi^A=A^\dagger\psi^S$, where $A=\sum_{\beta=L,R} |\beta (t)\ra\la \beta(0)|$
and $\psi^S$ is the Schr\"odinger-picture wave function,  
$\psi^A$ obeys $i\hbar\dot{\psi}^A=(H_A -K_A) \psi^A$, with 
$H_A=A^\dagger H A$, and $K_A=i\hbar A^\dagger \dot{A}$. 
For real $\la x|R(t)\ra$ and $\la x|L(t)\ra$,  the symmetry $\la x|R(t)\ra=\la -x|L(t)\ra$ makes $K_A=0$.     
  
We may invert Eq. (\ref{eigenstates_tm}) to write the bare states in terms of ground and excited states,
and get $\delta (t)$ from Eq. (\ref{H_tm}). The actual dynamics 
is approximated by   
identifying $|\psi^{\pm}_0(t)\ra$ and $E^{\pm}_0(t)$ with the instantaneous ground and excited states and energies of the unperturbed FF Hamiltonian.
They are combined to compute the bare basis in coordinate representation,  and with them the matrix elements $\la\beta'|H_{\lambda}|\beta\ra$.
The dynamics in the moving frame for the two-mode Hamiltonian may then be solved. 
%
%
%
%
%
%

{\it{Sudden approximation}.---}
The behaviour at low $\lambda$ may be understood with the sudden approximation \citep{Messiah}. Its validity
requires \citep{Messiah}
%
$
t_f \ll \hbar/\Delta \overline{H_A}, 
$
%
where $\Delta \overline{H_A}=\sqrt{\langle\psi(0)|\overline{H_A}^2|\psi(0)\rangle-\langle\psi(0)|\overline{H_A}|\psi(0)\rangle^2}$
and  $\overline{H_A}=\frac{1}{t_f}\int_{0}^{t_f}dt'H_A(t')$. 
With $|\psi(0)\rangle=|\psi_{0}^{-}(0)\rangle$
the condition to apply the sudden approximation becomes 
$\lambda \ll \frac{2\hbar}{t_f}$.  
In this regime the dynamical wave function $\psi(t_f)$ is not affected by the perturbation and becomes the ideal split state $\psi^-_0(t_f)$, up to a phase factor. 

The previous results may be extended to weakly 
non-linear Bose-Einstein condensates for which $g_1/a_0\lambda\ll1$.
Otherwise the instability 
of adiabatic splitting with respect to perturbations 
is strongly suppressed by the compensating
effect of the non-linear term \citep{split}. Of course the shortcuts would still
be useful if the time is to be reduced.   
\section{Discussion}
We have presented an overview of recent work  
on shorcuts to adiabaticity (STA) covering a 
broad span of methods and physical systems. 
STA offer many promissing
research and application avenues with practical and fundamental implications.  
Several pending tasks have been described along the text. We add here some more:   
To extend the set of basic physical limitations
and laws for fast processes in specific operations,
taking into account different constraints;   
To generate simple, viable shortcuts making systematic use of symmetries; 
To enhance  
robustness versus different types of noise and perturbations;  
To perform inverse-engineering with invariants beyond the quadratic-in-momentum family; 
To develop shortcuts for adiabatic computing, and in general for Hamiltonians that cannot be easily diagonalized, as in \cite{Cal11}; 
To design or supplement STA by 
optimal control theory methods. We have seen some exaples but many other optimization problems await unexplored.      
               
Indeed STA open interesting prospects to 
improve or make realizable quantum information and technology operations, by 
implementing new fast and robust transport or expansion approaches,  
internal state manipulations, and cooling protocols; nuclear magnetic resonance 
is another field where developing ideal pulses may benefit from STA.
STA could be also useful beyond single or many-body quantum systems, 
e.g. to build optical short-length mode converters or for 
designing mechanical operations with nanoparticles, mesoscopic, 
or macroscopic objects. In classical mechanics there are 
many examples of adiabatic evolution that may be shortcut. The application of this concept to interacting classical gas manipulation remains also an open question. 

We have witnessed in a few years a surge of activity 
and applications that could have hardly been predicted.  
Researchers creativity will likely continue to surprise us 
in the stimulating crossroad of STA with new, unnexpected
concepts and applications.

{\it Acknowledgment.---} 
We are grateful to D. Alonso, Y. Ban, M. Berry,  M. G. Boshier, B. Damski,  J. Garc\'ia-Ripoll, J.-S. Li, G. C. Hegerfeldt, R. Kosloff, M. A. Mart\'\i n-Delgado, I. Lizuain,  D. Porras, M. B. Plenio, M. Rams,  L. Santos, S. Schmidt, E. Sherman,  
D. Stefanatos, E. Timmermans, and W. H. Zurek.  

We acknowledge funding by Grants No. IT472-10, 
FIS2009-12773-C02-01, 61176118, NSF PHY11-25915, BFI09.39, BFI08.151, 12QH1400800
the UPV/EHU Program
UFI 11/55, the U.S Department of Energy through the LANL/LDRD Program, a  LANL J. Robert Oppenheimer fellowship (A.d.C.), and a UPV/EHU fellowship (S.M.G.). A.d.C. is grateful to KITP for hospitality.

\end{document}